\documentclass[%
reprint,
 amsmath,amssymb,
 aps,
pra,
floatfix,
longbibliography,
]{revtex4-2}

\usepackage{graphicx}
\usepackage{dcolumn}
\usepackage{bm}
\usepackage{lipsum}
\usepackage{xcolor}
\usepackage{multirow}
\usepackage{svg}
\usepackage{appendix}
\usepackage{mathtools}
\usepackage{makecell}
\usepackage{amsmath}
\usepackage{float}
\usepackage{accents}
\usepackage{braket}

\newcommand\authormark[1]{\textsuperscript{#1}}

\begin{document}

\preprint{APS/123-QED}
\title{{Photon triplets from integrated microrings:\\ A path towards deterministic non-Gaussianity on a chip}}

\author{Samuel E. Fontaine\authormark{1}, J. E. Sipe\authormark{1}, Marco Liscidini\authormark{2}, and Milica Bani\'c\authormark{3}}
\affiliation{\authormark{1}Department of Physics, University of Toronto, 60 St. George Street, Toronto, ON M5S 1A7, Canada\\ 
\authormark{2}Dipartimento di Fisica, Università di Pavia, Via Bassi 6, 27100 Pavia, Italy \\
\authormark{3}National Research Council of Canada, 100 Sussex Drive, Ottawa, Ontario K1A 0R6, Canada} 
\date{\today}

\begin{abstract}

We propose cascaded spontaneous four-wave mixing (SFWM) in microring resonators as a scalable and efficient approach for directly generating non-Gaussian states of light. Focusing on the well-understood ``low-gain" regime, we demonstrate that triplet generation through cascaded SFWM can be achieved with high efficiency and favorable spectral characteristics using realistic microring sources in AlGaAs. The ability to achieve the generation of light in a single set of supermodes -- and the  {predicted} accessibility of the ``high-gain" regime at realistic pump powers -- makes this source a promising candidate as a direct and deterministic source of non-Gaussian light for photonic quantum information processing.
\end{abstract}

\maketitle

\section{Introduction}
Generating non-Gaussian states of light is a central topic in the quantum optics and photonic quantum information processing communities. Due to their uniquely ``non-classical" characteristics, non-Gaussian states are a necessary resource for implementing photonic quantum information processing \cite{spekkens_PhysRevLett.101.020401}. But creating non-Gaussian states is challenging \cite{walschaers_PRXQuantum.2.030204}. The conventional approach is to apply measurements -- usually photon-number-resolving (PNR) detection -- to squeezed light \cite{Su_PhysRevA.100.052301, PhysRevA.110.012436,Larsen2025}. Although PNR detectors and squeezed light sources are readily available, relying on detection to introduce non-Gaussianity has the drawbacks of being probabilistic, introducing latency times, and requiring cryogenic components in a setting that could otherwise operate at room temperature. This has led to a search for \textit{deterministic}, room-temperature sources of non-Gaussian light based on the use of parametric processes \cite{PRXQuantum.4.010333}. 

Any such source must satisfy at least three requirements to be of broad practical interest. First, it must have a sufficiently high \emph{brightness}. Second, for the generated light to be useful and easily characterized it must be described by a \textit{small set of supermodes} \cite{arxiv:osorio2025}. Finally, \emph{scalability} -- for example, through on-chip integration -- is required for the source to be relevant in the large-scale information-processing applications that require non-Gaussianity as a resource.

For a parametric nonlinearity to induce non-Gaussianity, it must be described by a Hamiltonian that is at least cubic in bosonic operators \cite{Lloyd_PhysRevLett.82.1784}. One set of proposals involves the use of third-order parametric down-conversion (TOPDC), a non-Gaussian unitary process that can be described as a higher-order squeezing interaction in which photons are produced in sets of triplets. Many TOPDC-based sources have been proposed, yet this approach is hampered by the challenge of achieving phase matching, and the weakness of the $\chi^{(3)}$ nonlinearity on which the process relies. 

Another proposal relies on pump depletion and the resulting entanglement of pump light with photon pairs generated by the pump. This has been theoretically considered in the so-called mesoscopic regime in scenarios involving long waveguides \cite{Yanagimoto:22_mesoscopic_nG,yanagimoto2023quantum}, and in microring resonators requiring very high finesse and very high pump powers \cite{Vendromin25}. Experimental implementation of these schemes has been hampered by fabrication requirements and material limitations, and issues of scalability and few-supermode generation can also be of concern.

An alternative strategy would be the use of cascaded squeezing, where a squeezed vacuum is generated and then used to drive another squeezed light source; because the second process is driven by a non-classical pump, it has the desired non-Gaussian form \cite{Yanagimoto:22_mesoscopic_nG, PRXQuantum.4.010333}. If the first squeezing interaction is restricted to the low-gain regime, this results in the probabilistic generation of photon triplets: This process -- and the certification of the resulting tripartite entanglement -- has been implemented through spontaneous parametric down-conversion (SPDC) in bulk crystals \cite{Hbel2010_nature, Hamel2014, leger_PhysRevResearch.5.023131, chaisson_PhysRevA.105.063705} and in integrated waveguides \cite{Krapick:16}{, and in cascaded quantum dot schemes \cite{Khoshnegar2017PhotonTriplets}}. Besides the inherent scalability, the use of {fully} integrated structures offers an enhanced effective nonlinearity due to the strong transverse field confinement. 

Yet implementations of cascaded squeezing in microring resonators have not yet been investigated in detail, despite the fact that microring sources satisfy the three requirements mentioned above. First, they provide an additional resonant field enhancement, further increasing the structure's effectiveness in light generation and, importantly, increasing the brightness by having comparable or higher generation rates within a smaller bandwidth than non-resonant integrated sources \cite{Helt:12}. We will also show that the type of ``few-supermode" operation we seek can be obtained very effectively using resonant sources, often achieving a higher level of separability than non-resonant sources \cite{arxiv:osorio2025,houde2024ultrashortpulsepumpedsinglemodetype0squeezers}. Finally, the scalability of microring resonators has already made them a popular choice in the implementation of large-scale integrated devices.

The enhanced nonlinearity in these devices allows one to envision implementing cascaded \emph{spontaneous four-wave mixing} (SFWM) rather than SPDC. This has a number of practical advantages: The energy conservation conditions in SFWM allow for all the fields to be very close to each other in frequency, for instance in the same telecom band. Phase matching -- which can be a significant challenge in triplet generation -- is therefore widely accessible. The narrow frequency range should also make it relatively easy to find appropriate material platforms that are transparent at the relevant wavelengths and that do not exhibit deleterious effects such as two-photon absorption. 

For the sample calculations, we envision the use of integrated AlGaAs structures, but the scheme we propose can be implemented in other platforms. As well, to benchmark the results that can be expected in first experiments, we focus on the probabilistic generation of photon triplets. We will also show that one can go beyond that limit with reasonable experimental parameters.

\section{Cascaded spontaneous four-wave mixing}
\begin{figure}[t]
    \centering
\includegraphics[width=0.95\linewidth]{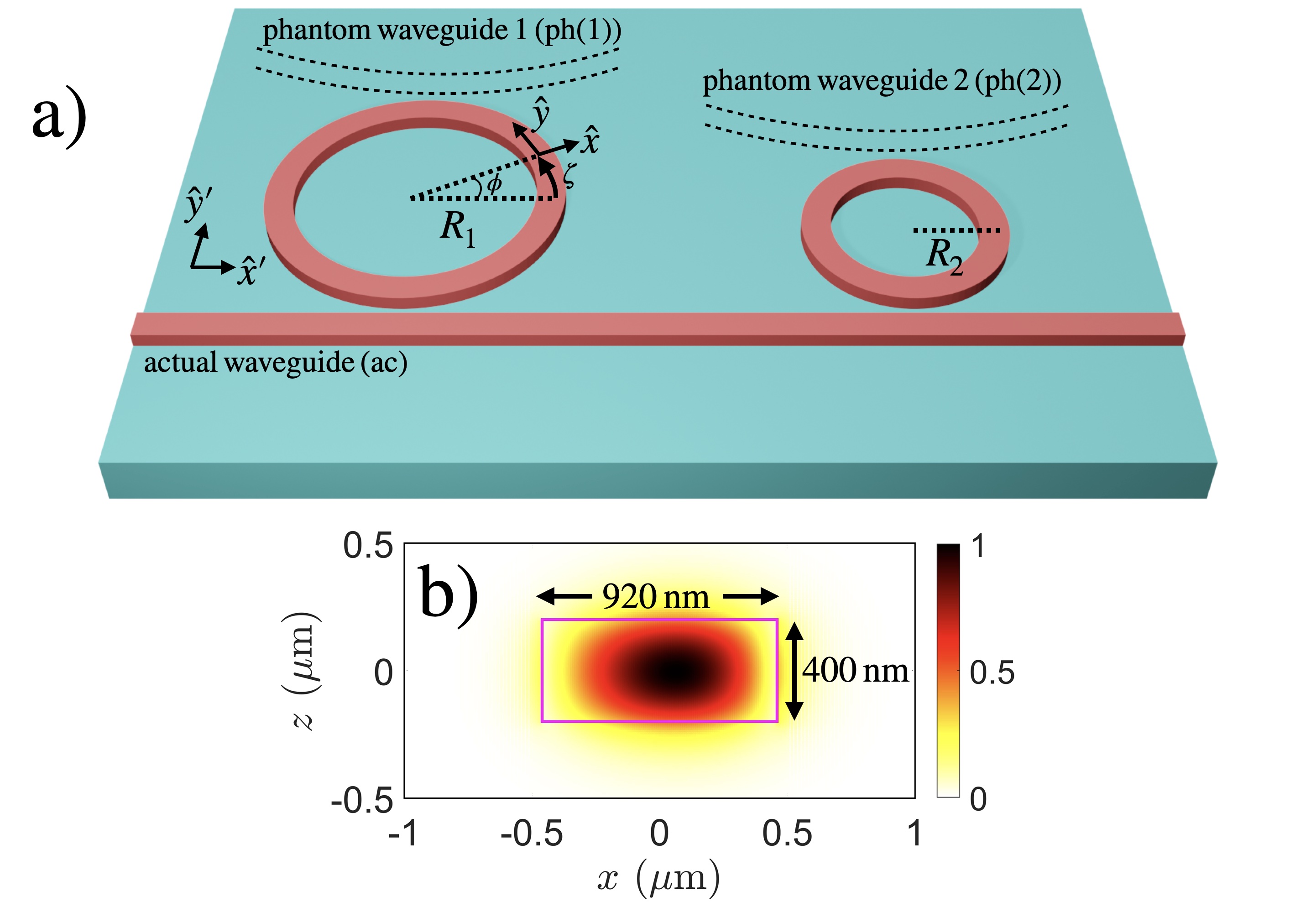}
    \caption{a) Subsequent microring resonators coupled to a bus waveguide, each coupled to a phantom waveguide to model photon losses. The $\hat{z}=\hat{z}'$ unit vectors are perpendicular to the plane of the chip. b) Electric field ($E_x$) mode profile of the TE$_{00}$ mode. Mode simulations are done in Lumerical with a bend radius of 10 $\mu\mathrm{m}$, using the refractive index for Al$_{0.3}$Ga$_{0.7}$As from Adachi et al. \cite{bib:Adachi1985}.}
    \label{fig:rings_modes}
\end{figure}

We propose a source consisting of two rings (see Fig. \ref{fig:rings_modes}a), such that the first (second) pair generation process occurs in the first (second) ring: This is ensured by tuning the positions of the ring resonances, resulting in a flexible and practical design in which each ring's resonance frequencies, free spectral range (FSR), and waveguide coupling can be controlled independently. Although single-ring cascaded SFWM could be implemented in principle, this would be challenging, and it would require careful dispersion engineering to ensure the proper resonance alignment. A device with two rings also allows for the two pump fields to be confined to separate rings, preventing some of the parasitic processes that would occur if both interactions occurred in a single ring. The basic microring structures sketched in Fig. \ref{fig:rings_modes}a are sufficient for demonstrations of cascaded SFWM, but in principle they could be replaced by more sophisticated structures -- e.g., with interferometric couplers or photonic molecules
-- to provide even more tunability of individual resonances, and suppression of unwanted nonlinear processes \cite{Vernon:17, menotti_PhysRevLett.122.013904, Zhang2021, seifoory_PhysRevA.105.033524}. 

\begin{figure}[h]
    \centering
\includegraphics[width=0.95\linewidth]{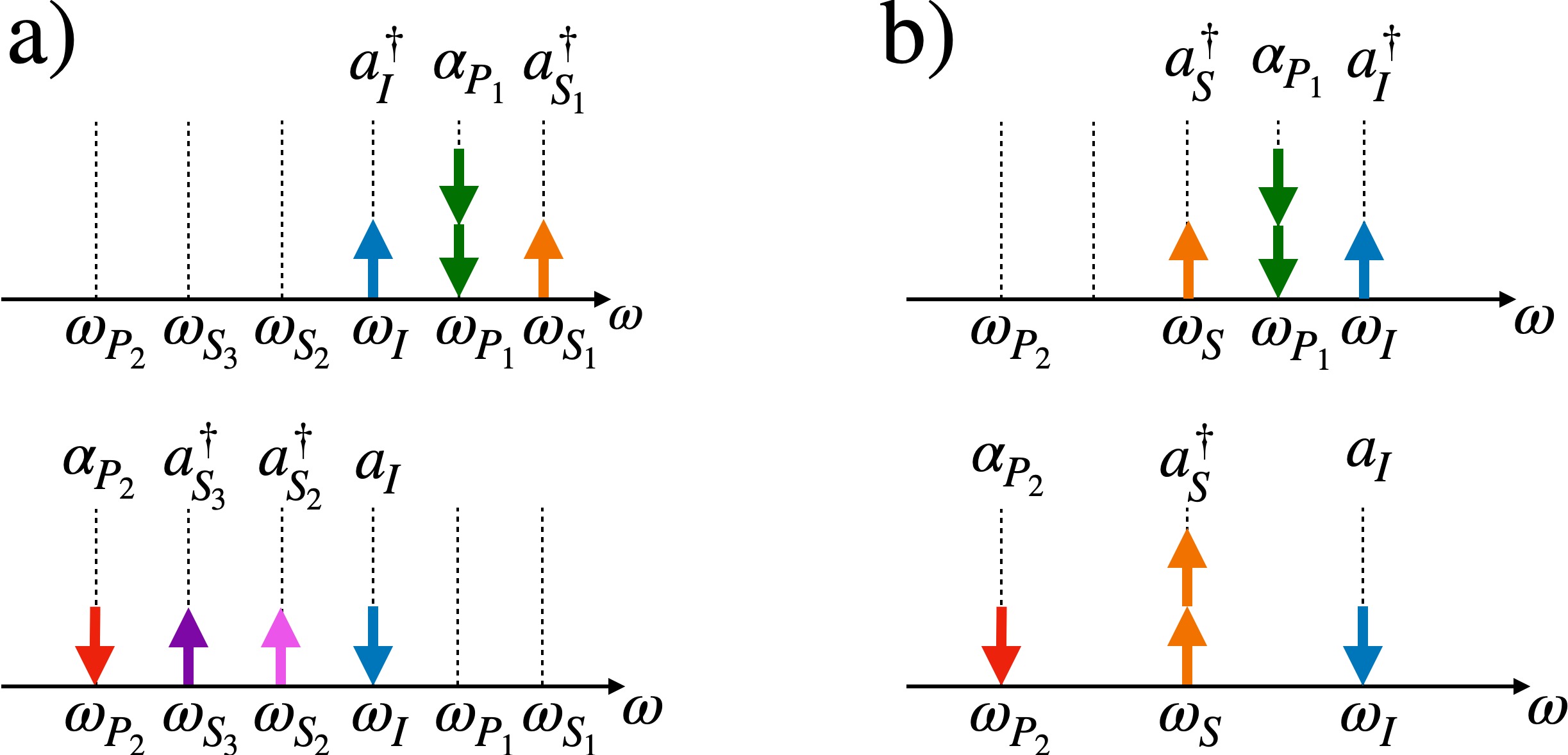}
    \caption{Energy diagrams for the generation of non-degenerate triplets (a), and degenerate triplets (b). The SFWM process for the first (second) ring is given in each subfigure's top (bottom) panel.}
    \label{fig:energyDiagrams}
\end{figure}

We envision working at telecom wavelengths, and we focus on a scheme in which the generated modes are non-degenerate in frequency (see Fig. \ref{fig:energyDiagrams}a). We choose this configuration to clearly identify the different modes involved in the interactions, but the design can be modified to affect the positions of the pump and generated fields; for example, degenerate triplet sources can be designed by adjusting the positions of the ring resonances (see Fig. \ref{fig:energyDiagrams}b).

The Hamiltonian describing the nonlinearity is (\cite{Quesada:22})
\begin{align}
    H_{\mathrm{NL}} = -\frac{1}{4\epsilon_0} \int \mathrm{d}\boldsymbol{r} \Gamma^{(3)}_{ijkl} (\boldsymbol{r})D^{i}(\boldsymbol{r}) D^{j}(\boldsymbol{r}) D^{k}(\boldsymbol{r}) D^{l}(\boldsymbol{r}),
\end{align}
where repeated Cartesian indices are implicitly summed over and where $\boldsymbol{D}(\boldsymbol{r})$ denotes the displacement field in the structure \cite{Quesada:22}, which we expand in terms of asymptotic fields \cite{Liscidini_asyfields, bethebreit_PhysRev.93.888} having the form 
\begin{align}
    \boldsymbol{D}_u^{\mathrm{in}}(\boldsymbol{r}) = \sum_{\lambda} \int \mathrm{d}k \boldsymbol{\mathcal{D}}_{uk}^{\mathrm{in}, \lambda}(\boldsymbol{r}) a_u^{\mathrm{in},\lambda}(k) + \mathrm{H.c.,} \label{eq:Dasy_def}
\end{align}
and similarly for the ``out" fields. We show the explicit expressions for the linear and nonlinear Hamiltonians in Appendices \ref{appendix:asyfields} and \ref{appendix:NL}, respectively. The tensor $\Gamma^{(3)}$ is related to the usual third-order nonlinear susceptibility tensor $\chi^{(3)}$ \cite{Quesada:22}, as presented in Appendix C (see Eq. \eqref{eq:Tensors}). In Eq. \eqref{eq:Dasy_def}, the index $\lambda$ denotes an input or output channel through which light can propagate \cite{PhysRevA.106.043707, PhysRevA.111.023705}, and for the structure given in Fig. \ref{fig:rings_modes}a, $\lambda\in\{\mathrm{ac,ph(1),ph(2)}\}$ for the actual, first, and second phantom channels; the phantom channels are used to account for scattering losses \cite{Liscidini_asyfields}. The $a_u^{\text{in/out},\lambda}(k)$ denote bosonic lowering operators with the usual commutation relations \cite{PhysRevA.106.043707}, and $\boldsymbol{\mathcal{D}}_{uk}^{\text{in/out},\lambda}(\boldsymbol{r})$ denotes the asymptotic field amplitude. The form of the latter depends on the structure in question, and we detail those for the cascaded rings from Fig. \ref{fig:rings_modes}a in Appendix \ref{supp:Fields}.

We restrict ourselves to the terms in the full nonlinear Hamiltonian relevant to cascaded SFWM: $H_{\mathrm{NL}}^{(I)}(t) = H_{\mathrm{SFWM1}}^{(I)}(t) + H_{\mathrm{SFWM2}}^{(I)}(t)$ (see Appendix \ref{appendix:NL}). The state generated by the nonlinear interactions is given by
\begin{align}
    \ket{\psi(t)} = \mathcal{T} \text{exp} \left(-\frac{i}{\hbar} \int_{t_0}^{t} \mathrm{d}t' H_{\mathrm{NL}}^{(I)}(t')\right) \ket{\text{vac}}, \label{eq:SE}
\end{align}
where $\mathcal{T}$ is the time ordering operator. In this manuscript, we address the typical ``low-gain" regime for the {cascaded processes} \cite{Hamel2014, Krapick:16}, such that Eq. \eqref{eq:SE} can be evaluated perturbatively: $\ket{\psi}\approx\ket{\mathrm{vac}}+\beta\ket{\mathrm{II}}+\sigma\ket{\mathrm{III}}$, where both $\beta$ and $\sigma$ are small parameters, and $\ket{\mathrm{II}}$ and $\ket{\mathrm{III}}$ are two- and three-photon states respectively. For the generated state, we obtain an expression of the form 
\begin{align}
    \ket{\psi(\infty)} & \approx \ket{\text{vac}} + \beta \sum_{\lambda\lambda'}\int \mathrm{d}k_1 \mathrm{d}k_2 \varphi^{\lambda \lambda'}(k_1,k_2;t\rightarrow\infty) \nonumber \\
    &  \hspace {2.5cm}\times \left(a^{\lambda}_{S_{1}}(k_1) a^{\lambda'}_{I}(k_2) \right)^{\dagger}\ket{\text{vac}} \nonumber \\ 
    &+ \sigma \sum_{\mu\mu'\mu''}\int \mathrm{d}k_1 \mathrm{d}k_2 \mathrm{d}k_3 \Psi^{\mu \mu' \mu''}(k_1,k_2,k_3;t\rightarrow \infty)\nonumber \\
    & \hspace{0.5cm}\times \left(a_{S_{1}}^{\mu}(k_1) a_{S_{2}}^{\mu'}(k_2) a_{S_{3}}^{\mu''}(k_3)\right)^{\dagger} \ket{\text{vac}} + \dots, \label{eq:ket}
\end{align}
where we have dropped the ``out" label from the operators. The second term in Eq. \eqref{eq:ket} refers to the generation of a photon pair in the first SFWM process. As described earlier \cite{PhysRevA.106.043707}, $\varphi^{\lambda \lambda'}(k_1,k_2;t)$ denotes the biphoton wavefunction (BWF) that describes the correlations between a pair of photons that couple out of the device via the channels $\lambda$ and $\lambda'$, and is given in Eq. \eqref{eq:BWF_def}. 

The final term in Eq. \eqref{eq:ket} refers to the generation of a photon triplet through cascaded SFWM. The spectral correlations for a photon triplet are characterized by the \textit{triphoton} wavefunction (TWF), given in Eq. \eqref{eq:TWF_def}. We note that this triplet term can only arise due to the non-Gaussian nature of the cascaded SFWM Hamilitonian (Eq. \eqref{eq:HSFWM2}); therefore, a demonstration of triplet generation can be considered a signature of the state's non-Gaussianity.

The probabilities per pump pulse of generating pairs from the first process and triplets from the cascaded processes are $|\beta|^2$ and $|\sigma|^2$, respectively: These are computed by imposing the normalization conditions in Eqs. \eqref{eq:BWF_full_norm}--\eqref{eq:TWF_full_norm} on the BWF and TWF. A corresponding probability for photon pairs coupling out through channels $\lambda$ and $\lambda'$ can be denoted $|\beta^{\lambda\lambda'}|^2$; these quantities can be obtained from the BWF as described earlier \cite{PhysRevA.111.023705}. Similarly, one can also compute specific probabilities $|\sigma^{\mu\mu'\mu''}|^2$ to evaluate, for example, the rate of threefold coincidences in the physical output channel. The state of interest for triphoton detection is the one where all three photons are at the physical output channel (i.e. $\mu=\mu'=\mu''=\mathrm{ac}$), and is characterized by $\Psi^{\mathrm{ac,ac,ac}}(k_1,k_2,k_3;t\rightarrow \infty)$. This state can be written as
\begin{align}
\ket{\mathrm{III}'}& \equiv \mathcal{N} \int \mathrm{d}k_1 \mathrm{d}k_2 \mathrm{d}k_3 \Psi^{\mathrm{ac,ac,ac}}(k_1,k_2,k_3;t\rightarrow \infty)\nonumber \\
    & \hspace{1cm}\times \left(a_{S_{1}}^{\mathrm{ac}}(k_1) a_{S_{2}}^{\mathrm{ac}}(k_2) a_{S_{3}}^{\mathrm{ac}}(k_3)\right)^{\dagger} \ket{\text{vac}},
    \label{eq:ac_ac_ac}
\end{align}
where $\mathcal{N}$ is a normalization constant {not equal to unity because one must sum over all the output channels (Eq. \eqref{eq:TWF_full_norm}) for the wavefunction $\Psi^{\mu\mu'\mu''}(k_1,k_2,k_3;t\rightarrow \infty)$ to be properly normalized.} When referring to the generated state of triplets for the remainder of this manuscript, we will be considering the experimentally accessible state $\ket{\mathrm{III}'}$. 

\section{Few-supermode operation}
The ability to generate light in the minimal number of supermodes is essential in developing a useful source of non-Gaussianity, just as one seeks single-Schmidt-mode sources of squeezed light \cite{Zielnicki2018, arxiv:osorio2025,Knill2001,PhysRevLett.119.170501,bulmer2025simulatinglossypartiallydistinguishable}. In the low-gain regime this is achieved by obtaining a TWF that is separable \cite{arxiv:osorio2025}, and although more work is needed to understand non-Gaussian nonlinear optics in the high gain regime, this approach can be expected to be useful there as well. 

The separability of the TWF can be characterized by referring to the purities of the reduced density matrices
\begin{align}
\rho^{(1)}(k_1,k_1') &= |\mathcal{N}|^2\int \mathrm{d}k_2\mathrm{d}k_3 \Psi^{\mathrm{ac,ac,ac}}(k_1,k_2,k_3;t\rightarrow\infty) \nonumber \\
& \hspace{1.2cm}\times \left(\Psi^{\mathrm{ac,ac,ac}}(k_1',k_2,k_3;t\rightarrow\infty)\right)^*
    \label{eq:reduced}
\end{align}
(and similarly for $\rho^{(2)}$ and $\rho^{(3)}$) \cite{arxiv:osorio2025}. In the limit of a perfectly separable TWF, implying the light is generated in a single set of supermodes, $p_1 \equiv \mathrm{Tr}\left(\rho^{(1)}(k_1,k_1')\right)=1$. Thus, to design a source with our desired mode structure, we seek a configuration in which $p_1$, $p_2$, and $p_3$ are as close as possible to unity. In other contexts, it is known that one can shape the joint spectral amplitude produced by a resonant device by manipulating the pump durations \cite{Shukhin:24,PhysRevA.111.023705}; in the following sample calculations, we show that this strategy also applies to the TWF in cascaded SFWM.

\section{Sample calculation}
We now compute triplet generation TWFs and rates for a realistic device. We consider implementations with AlGaAs structures: This emerging platform supports large nonlinear coefficients $\gamma_{\mathrm{NL}}$, has demonstrated high quality factors \cite{steiner_PRXQuantum.2.010337}, and is compatible with hybrid integration approaches \cite{Baboux:23}. In our calculations, we take $\chi^{(3)} \approx 5.6\times10^{-19}$ m$^2$/V$^2$ ($n_2 \approx 1.43\times10^{-17}$ m$^2$/W) \cite{Aitchison1997,steiner_PRXQuantum.2.010337, Baboux:23}, corresponding to nonlinear parameters of $\gamma_{\mathrm{NL}}\approx250\,(\mathrm{W\cdot m})^{-1}$ for both SFWM processes, computed using the TE$_{00}$ modes (see Fig. \ref{fig:rings_modes}b and {Appendix \ref{Supp}}). For the ring dimensions we take $R_1 = {20\,\mu\mathrm{m}}$ and $R_2 = {10\,\mu\mathrm{m}}$ (recall the labelling in Fig. \ref{fig:rings_modes}a), {both with cross-sectional dimensions of $(400 \times 920)$ nm (see Fig. \ref{fig:rings_modes}b)}. We assume {realistic intrinsic quality factors of $Q_{\mathrm{int}} = 10^6$ for all the resonances in the frequency range of interest} \cite{steiner_PRXQuantum.2.010337}. The rings and pumps should be tuned such that both pumps are on resonance. 

We consider the pump fields $P_{1}$ and $P_{2}$ for the first and second ring respectively (recall Fig. \ref{fig:rings_modes} \& \ref{fig:energyDiagrams}), to be centered {at wavelengths} $\lambda_{P_{1}}=1546.83$ nm and ${\lambda_{P_{2}}=1584.29}$ nm, respectively. The triplets are generated in the resonances with center wavelengths $\lambda_{S_{1}}=1541.60$ nm, $\lambda_{S_{2}}=1562.68$ nm, and $\lambda_{S_{3}}=1573.41$ nm. The rate of triplets at the chip's output is given by $R_3=R_R\eta_{S_{1}}^{\mathrm{ac}(1)}\eta_{S_{2}}^{\mathrm{ac}(2)}\eta_{S_{3}}^{\mathrm{ac}(2)}|\sigma|^2$, where $R_R$ is the pulse repetition rate, and $\eta_{u}^{\lambda(i)}$ are the escape efficiencies (see Eq. \eqref{eq:EscEff}). 

To ensure the validity of our perturbative treatment (recall Eq. \eqref{eq:ket}), we limit the average power in the first pump $P_{1}$ to yield a pair generation probability of $|\beta|^2\approx 0.1$. The power in the second pump $P_{2}$ will be set to avoid any material damage due to linear and nonlinear absorption \cite{May:19,10.1063/1.3119629,PRXQuantum.6.010338}. The pump durations and relative escape efficiencies impact the triphoton wavefunctions and rates; we show this by considering three different pumping configurations, summarized in Tab. \ref{table1}.
\begin{table*}[]
\begin{tabular}{c||ccc}
                                              & \multicolumn{3}{c}{Configurations of first pump ($P_{1}$) / second pump ($P_{2}$)}                     \\ \hline \hline
{Configuration}                               & \multicolumn{1}{c|}{($A$)}                   & \multicolumn{1}{c||}{($B$)}                   & ($C$)          \\ \hline 
{Duration (FWHM)}                             & \multicolumn{1}{c|}{300 ps / CW}             & \multicolumn{1}{c||}{300 ps / 300 ps}         & 50 ps / 100 ps   \\ \hline 
{Pulse energy}                                & \multicolumn{1}{c|}{0.17 pJ / NA}            & \multicolumn{1}{c||}{0.17 pJ / 36.7 pJ}       & 0.52 pJ / 20.5 pJ     \\ \hline
{Repetition rate}                             & \multicolumn{1}{c|}{10 MHz / NA}             & \multicolumn{1}{c||}{10 MHz / 10 MHz}         & 10 MHz / 10 MHz  \\ \hline
{Peak power in waveguide}                     & \multicolumn{1}{c|}{0.53 mW / 1.4 mW}        & \multicolumn{1}{c||}{0.53 mW / 0.12 W}        & 9.74 mW / 0.19 W      \\ \hline 
{Average power in waveguide}                  & \multicolumn{1}{c|}{1.60 $\mu$W / 1.4 mW}    & \multicolumn{1}{c||}{1.60 $\mu$W / 0.34 mW}      & {4.87 $\mu$W / 0.19 mW}   \\ \hline \hline
{Purity $p_1$ / $p_2$ / $p_3$}                & \multicolumn{1}{c|}{0.974 / 0.610 / 0.610}   & \multicolumn{1}{c||}{0.982 / 0.806 / 0.806}   & 0.995 / 0.970 / 0.970  \\ \hline
{$|\sigma|^2$}                                & \multicolumn{1}{c|}{5.94$\times10^{-6}$}     & \multicolumn{1}{c||}{1.54$\times10^{-5}$}     & 2.52$\times10^{-6}$ \\ \hline
{Rate}                                        & \multicolumn{1}{c|}{7.43 Hz}                 & \multicolumn{1}{c||}{19.2 Hz }                & 5.66 Hz          
\end{tabular}
\caption{Simulation results for different pump configurations, with $Q_{\mathrm{int}}=10^6$. Critical coupling ($\eta=0.5$) is used for configurations $A$ and $B$, and $\eta_{P_{1}}^{\mathrm{ac}(1)}=0.95$, $\eta_{S_{1}}^{\mathrm{ac}(1)}=\eta_{I}^{\mathrm{ac}(1)}=0.9$, $\eta_{P_{2}}^{\mathrm{ac}(2)}=\eta_{I}^{\mathrm{ac}(2)}=0.85$, and $\eta_{S_{2}}^{\mathrm{ac}(2)}=\eta_{S_{3}}^{\mathrm{ac}(2)}=0.5$ is used for configuration $C$.}
\label{table1}
\end{table*}

Initially we consider two configurations $A$ and $B$, with critical coupling for all resonances, such that all loaded quality factors are $5\times10^{5}$. We take $P_{1}$ to be a Gaussian pulse with full-width-half-maximum (FWHM) of $300$ ps -- similar to the ring dwelling time -- with a repetition rate $R_R=10$ MHz. {Interesting triplet rates can be obtained with either pulsed or CW excitation in the second ring. }In configuration $A$, $P_2$ is CW; in configuration $B$, it is a train of $300$ ps pulses. To obtain $|\beta|^2\approx 0.1$, we limit $P_{1}$ to a peak power of $0.53\,\mathrm{mW}$ in the bus waveguide. In configuration $A$, we take the power of the CW $P_{2}$ to be ${1.4}$ mW, which corresponds to $\sim1$ W circulating in the ring; this results in an output triplet rate of $R_3 \approx {7.43}$ Hz. In configuration $B$, the pulsed $P_2$ is set such that the peak power in the ring does not exceed 10 W; this results in a peak power in the waveguide of 0.12 W, giving $R_3 = 19.2$ Hz. 

\begin{figure}[h]
    \centering
\includegraphics[width=1\linewidth]{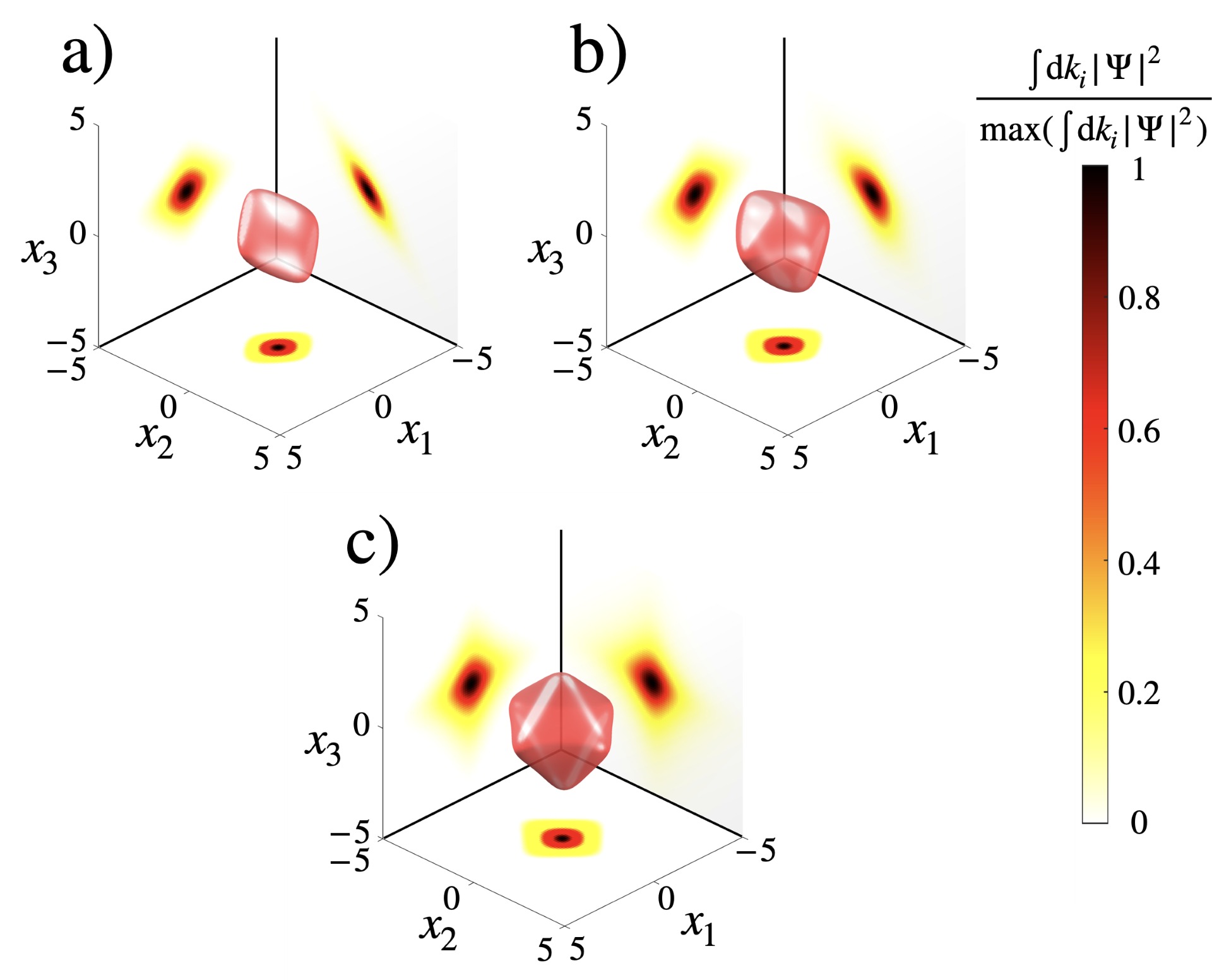}
    \caption{
    {Visualization of the total triphoton wavefunction $|\Psi|^2\equiv|\Psi^{\mathrm{ac,ac,ac}}(k_1,k_2,k_3;t\rightarrow\infty)|^2$ for configuration $A$ (a), $B$ (b) and $C$ (c); we use $x_i = v_{S_i}(k_i-K_{S_i})/\bar{\Gamma}_{S_i}$,} where $K_{S_i}$ indicates the reference wavenumber for the $i^{th}$ signal, and in the panel perpendicular to the $x_i$ axis the quantity indicated above the color bar is plotted using the scale of the bar. }
    \label{fig:TWFs}
\end{figure}

We represent the TWFs obtained with configurations $A$ and $B$ in Fig. \ref{fig:TWFs}a and \ref{fig:TWFs}b, respectively. In moving from configuration $A$ to configuration $B$, we see -- similar to photon-pair generation in a ring resonator \cite{Helt:10_SFWMringresonator} -- that pumping with a pulse duration shorter than the dwelling time in the ring results in a TWF with reduced spectral correlations. The visualizations of the TWFs in Fig. \ref{fig:TWFs} are isosurfaces at 10\% of the modulus squared of the maximal value of the given TWF, and projections over each variable are provided on the panels. {The projections on the panels of Fig. \ref{fig:TWFs} can be thought of as the spectral correlations between the two photons over which the TWF has not been integrated. For example, comparing the top right panels from Fig. \ref{fig:TWFs}a-c, we see a change in the spectral correlations between the photons in $S_2$ and $S_3$ (the perpendicular axis being the $S_1$ variable,  over which the integration is performed), going from more correlated to less correlated, which can be seen by the relative thickness of the off-diagonal (similar to what is seen in photon pair generation \cite{drago_PhysRevA.110.023710}). Similarly, the overall shape of the three-dimensional isosurfaces can describe the correlations between the three photons constituting the triplet.}

The spectral correlations can be further reduced by adjusting the coupling efficiencies together with the pump pulse durations \cite{Vernon:17}. In a third scenario, configuration $C$, we vary the coupling efficiencies, which is achievable by using interferometric couplers on the device \cite{tuneableQ_10930539}. The various coupling efficiencies are adjusted such that the loaded quality factors in the first ring are $5\times10^4$ for $P_{1}$, and $10^5$ for $S_{1}$ and $I$; in the second ring the coupling efficiencies are adjusted such that the loaded quality factors are $1.5\times10^5$ for $P_{2}$ and $I$, and $5\times10^5$ for $S_{2}$ and $S_{3}$. We adjust the pulse durations correspondingly, to be slightly shorter than the corresponding dwelling times for the various resonances. We take durations of $50$ ps for $P_{1}$ and $100$ ps for $P_{2}$, with peak powers in the waveguides of 9.74 mW (0.52 pJ pulse energy, and 4.87 $\mu$W average power) and 0.19 W (20.5 pJ pulse energy, and 0.19 mW average power), respectively. In this configuration, we obtain $|\sigma|^2 = 2.52\times10^{-6}$ and $R_3=5.66$ Hz. It is particularly interesting because it yields a highly uncorrelated TWF {(see Fig. \ref{fig:TWFs}c)}, with $p_1 = 0.995$, $p_2 = p_3 = 0.970$; for reference, a recent work on direct triplet generation in non-resonant sources quoted $p_1 = p_2 = p_3 \approx 0.8$ as the maximum purities that can be obtained through basic dispersion engineering \cite{arxiv:osorio2025}, which suggests that the purities achievable in such resonant systems can far surpass purities of other proposals and implementations in waveguides or other non-resonant sources \cite{houde2024ultrashortpulsepumpedsinglemodetype0squeezers,Vernon:17}.

We note that higher generation rates, brightnesses and purities can be achieved by tuning various pumping and coupling parameters. In this manuscript, our focus is on showing that high generation rates are possible for a range of purities. We leave the determination of the maximal generation rates and optimal purities to future work.

\section{Discussion}
Our proposed device demonstrates that photon triplet generation can be achieved on a platform that is simultaneously efficient, compact, and readily manufacturable using existing fabrication techniques. Our proposed design requires lower average pump powers compared to other proposals, requiring only average on-chip powers of $\sim$ 1 mW or less, and we have used well-accepted material parameters. Furthermore, it offers both spectral tunability and control over the number of supermodes, providing essential flexibility for tailoring the properties of the generated non-Gaussian states. It would yield triplet rates comparable to those quoted in other theoretical proposals \cite{cavanna_PhysRevA.101.033840, Moebius:16, Vernay:21, banicTOPDC_PhysRevA.106.013710}, while being more efficient, compact, and allowing for tunability of the light's spectral properties. For comparison to experimentally demonstrated sources, taking a loss of 3 dB per output channel \cite{Castro2022,pu2022design} for a total loss of 9 dB; this would result in a measured triplet coincidence rate of $\sim 1$ Hz, higher than {or similar to} the rates observed in experimental implementations of cascaded pair generation \cite{Hbel2010_nature,Hamel2014, Krapick:16, leger_PhysRevResearch.5.023131} {($<$ 1 Hz) and cascaded quantum dots \cite{ Khoshnegar2017PhotonTriplets} ($\sim$1 Hz).} In addition, we note that the proposed device could achieve higher triplet rates by further increasing the repetition rate, since the resonator dwelling times (maximally $\sim 400$ ps for critical coupling) do not limit the repetition rate to the 10 MHz used in our calculations. We believe these high rates and promising characteristics strongly motivate the fabrication and demonstration of these sources, including the characterization of the generated {non-Gaussian} light, which could be done through stimulated emission tomography \cite{DominguezSerna2020}, for example. {In the low-gain regime considered here, genuine three-photon coincidences would constitute sufficient evidence that the emitted light possesses non-Gaussian features.}

Perhaps most interesting is the fact that this source, due to its high efficiency, is a strong candidate for implementing cascaded SFWM in a \emph{high-gain} regime: Here, we restricted the first SFWM process to a low-gain regime by restricting the energy of $P_1$ to less than a picojoule. From an experimental perspective, one can easily envision increasing the strength of $P_1$, resulting in squeezed vacuum which would then drive the second SFWM process, where in this high-gain regime measurements of quadrature non-Gaussian features, such as Wigner negativity, in phase space \cite{walschaers_PRXQuantum.2.030204} can be envisioned. The dynamics of cascaded SFWM in this high-gain regime, and the characteristics of the deterministically generated non-Gaussian state it would produce, will be a rich and challenging topic for future theoretical work. It will also be a novel source of non-classical light in more complex quantum photonic devices \cite{PRXQuantum.4.010333,PhysRevApplied.23.014003, crescimanna_PhysRevA.109.023717}.

\begin{acknowledgments}
S.E.F. and J.E.S. acknowledge the Natural Sciences and Engineering Research Council of Canada{, and the} European Union’s Horizon Europe Research and Innovation
Programme (101070700, project MIRAQLS) for financial support, and thank Colin Vendromin for insightful discussions. S.E.F. acknowledges support from a Walter C. Sumner Memorial Fellowship. M.L. acknowledges support by PNRR MUR Project No. PE0000023-NQSTI. M.B. acknowledges support from the Quantum Research and Development Initiative, led by the National Research Council Canada, under the National Quantum Strategy, and thanks Nicol\'as Quesada and Boris Braverman for stimulating discussions. \end{acknowledgments}

\bibliography{apssamp}

@article{Khoshnegar2017PhotonTriplets,
  title        = {A solid state source of photon triplets based on quantum dot molecules},
  author       = {Khoshnegar, Milad and Huber, Tobias and Predojevi{\'c}, Ana and Dalacu, Dan and Prilm{\"u}ller, Maximilian and Lapointe, Jean and Wu, Xiaohua and Tamarat, Philippe and Lounis, Brahim and Poole, Philip and Weihs, Gregor and Majedi, Hamed},
  journal      = {Nature Communications},
  volume       = {8},
  pages        = {15716},
  year         = {2017},
  doi          = {10.1038/ncomms15716},
  url          = {https://www.nature.com/articles/ncomms15716}
}

@article{Liscidini_asyfields,
  title = {Asymptotic fields for a Hamiltonian treatment of nonlinear electromagnetic phenomena},
  author = {Liscidini, M. and Helt, L. G. and Sipe, J. E.},
  journal = {Phys. Rev. A},
  volume = {85},
  issue = {1},
  pages = {013833},
  numpages = {15},
  year = {2012},
  month = {Jan},
  publisher = {American Physical Society},
  doi = {10.1103/PhysRevA.85.013833},
  url = {https://link.aps.org/doi/10.1103/PhysRevA.85.013833}
}

@article{arxiv:osorio2025,
doi = {10.1088/2058-9565/ae0759},
url = {https://doi.org/10.1088/2058-9565/ae0759},
year = {2025},
month = {sep},
publisher = {IOP Publishing},
volume = {10},
number = {4},
pages = {045045},
author = {Lorena Osorio, Gisell and Banić, Milica and Quesada, Nicolás},
title = {Strategies for generating separable photon triplets in waveguides and ring resonators},
journal = {Quantum Science and Technology}
}

@article{chaisson_PhysRevA.105.063705,
  title = {Phase-stable source of high-quality three-photon polarization entanglement by cascaded down-conversion},
  author = {Chaisson, Zachary M. E. and Poitras, Patrick F. and Richard, Mica\"el and Castonguay-Page, Yannick and Glinel, Paul-Henry and Landry, V\'eronique and Hamel, Deny R.},
  journal = {Phys. Rev. A},
  volume = {105},
  issue = {6},
  pages = {063705},
  numpages = {6},
  year = {2022},
  month = {Jun},
  publisher = {American Physical Society},
  doi = {10.1103/PhysRevA.105.063705},
  url = {https://link.aps.org/doi/10.1103/PhysRevA.105.063705}
}

@article{walschaers_PRXQuantum.2.030204,
  title = {Non-Gaussian Quantum States and Where to Find Them},
  author = {Walschaers, Mattia},
  journal = {PRX Quantum},
  volume = {2},
  issue = {3},
  pages = {030204},
  numpages = {68},
  year = {2021},
  month = {Sep},
  publisher = {American Physical Society},
  doi = {10.1103/PRXQuantum.2.030204},
  url = {https://link.aps.org/doi/10.1103/PRXQuantum.2.030204}
}

@article{spekkens_PhysRevLett.101.020401,
  title = {Negativity and Contextuality are Equivalent Notions of Nonclassicality},
  author = {Spekkens, Robert W.},
  journal = {Phys. Rev. Lett.},
  volume = {101},
  issue = {2},
  pages = {020401},
  numpages = {4},
  year = {2008},
  month = {Jul},
  publisher = {American Physical Society},
  doi = {10.1103/PhysRevLett.101.020401},
  url = {https://link.aps.org/doi/10.1103/PhysRevLett.101.020401}
}

@article{PhysRevA.110.012436,
  title = {Generation of flying logical qubits using generalized photon subtraction with adaptive Gaussian operations},
  author = {Takase, Kan and Hanamura, Fumiya and Nagayoshi, Hironari and Bourassa, J. Eli and Alexander, Rafael N. and Kawasaki, Akito and Asavanant, Warit and Endo, Mamoru and Furusawa, Akira},
  journal = {Phys. Rev. A},
  volume = {110},
  issue = {1},
  pages = {012436},
  numpages = {10},
  year = {2024},
  month = {Jul},
  publisher = {American Physical Society},
  doi = {10.1103/PhysRevA.110.012436},
  url = {https://link.aps.org/doi/10.1103/PhysRevA.110.012436}
}

@article{Larsen2025,
  title = {Integrated photonic source of Gottesman–Kitaev–Preskill qubits},
  volume = {642},
  ISSN = {1476-4687},
  url = {http://dx.doi.org/10.1038/s41586-025-09044-5},
  DOI = {10.1038/s41586-025-09044-5},
  number = {8068},
  journal = {Nature},
  publisher = {Springer Science and Business Media LLC},
  author = {Larsen,  M. V. and Bourassa,  J. E. and Kocsis,  S. and Tasker,  J. F. and Chadwick,  R. S. and González-Arciniegas,  C. and Hastrup,  J. and Lopetegui-González,  C. E. and Miatto,  F. M. and Motamedi,  A. and Noro,  R. and Roeland,  G. and Baby,  R. and Chen,  H. and Contu,  P. and Di Luch,  I. and Drago,  C. and Giesbrecht,  M. and Grainge,  T. and Krasnokutska,  I. and Menotti,  M. and Morrison,  B. and Puviraj,  C. and Rezaei Shad,  K. and Hussain,  B. and McMahon,  J. and Ortmann,  J. E. and Collins,  M. J. and Ma,  C. and Phillips,  D. S. and Seymour,  M. and Tang,  Q. Y. and Yang,  B. and Vernon,  Z. and Alexander,  R. N. and Mahler,  D. H.},
  year = {2025},
  month = jun,
  pages = {587–591}
}

@article{PRXQuantum.4.010333,
  title = {Quantum Nondemolition Measurements with Optical Parametric Amplifiers for Ultrafast Universal Quantum Information Processing},
  author = {Yanagimoto, Ryotatsu and Nehra, Rajveer and Hamerly, Ryan and Ng, Edwin and Marandi, Alireza and Mabuchi, Hideo},
  journal = {PRX Quantum},
  volume = {4},
  issue = {1},
  pages = {010333},
  numpages = {15},
  year = {2023},
  month = {Mar},
  publisher = {American Physical Society},
  doi = {10.1103/PRXQuantum.4.010333},
  url = {https://link.aps.org/doi/10.1103/PRXQuantum.4.010333}
}

@article{cavanna_PhysRevA.101.033840,
  title = {Progress toward third-order parametric down-conversion in optical fibers},
  author = {Cavanna, Andrea and Hammer, Jonas and Okoth, Cameron and Ortiz-Ricardo, Erasto and Cruz-Ramirez, Hector and Garay-Palmett, Karina and U'Ren, Alfred B. and Frosz, Michael H. and Jiang, Xin and Joly, Nicolas Y. and Chekhova, Maria V.},
  journal = {Phys. Rev. A},
  volume = {101},
  issue = {3},
  pages = {033840},
  numpages = {10},
  year = {2020},
  month = {Mar},
  publisher = {American Physical Society},
  doi = {10.1103/PhysRevA.101.033840},
  url = {https://link.aps.org/doi/10.1103/PhysRevA.101.033840}
}

@article{Vernay:21,
author = {Augustin Vernay and V\'{e}ronique Boutou and Corinne F\'{e}lix and David Jegouso and Florent Bassignot and Mathieu Chauvet and Benoit Boulanger},
journal = {Opt. Express},
number = {14},
pages = {22266--22274},
publisher = {Optica Publishing Group},
title = {Birefringence phase-matched direct third-harmonic generation in a ridge optical waveguide based on a KTiOPO4 single crystal},
volume = {29},
month = {Jul},
year = {2021},
url = {https://opg.optica.org/oe/abstract.cfm?URI=oe-29-14-22266},
doi = {10.1364/OE.432636},
}

@article{houde2024ultrashortpulsepumpedsinglemodetype0squeezers,
author = {Martin Houde and Liam Beaudoin and Robert Kwolek and Kazuki Hirota and Rajveer Nehra and Nicol\'{a}s Quesada},
journal = {Optica Quantum},
number = {6},
pages = {560--568},
publisher = {Optica Publishing Group},
title = {Ultrashort-pulse-pumped, single-mode type-0 squeezers in lithium niobate nanophotonics},
volume = {3},
month = {Dec},
year = {2025},
url = {https://opg.optica.org/opticaq/abstract.cfm?URI=opticaq-3-6-560},
doi = {10.1364/OPTICAQ.566703},
}

@book{bib:Boyd,
author = {Boyd, R. W.},
title = {Nonlinear Optics, Third Edition},
year = {2008},
isbn = {0123694701},
publisher = {Academic Press, Inc.},
address = {USA},
edition = {3rd}
}

@article{Zhang2021,
  title = {Squeezed light from a nanophotonic molecule},
  volume = {12},
  ISSN = {2041-1723},
  url = {http://dx.doi.org/10.1038/s41467-021-22540-2},
  DOI = {10.1038/s41467-021-22540-2},
  number = {1},
  journal = {Nature Communications},
  publisher = {Springer Science and Business Media LLC},
  author = {Zhang,  Y. and Menotti,  M. and Tan,  K. and Vaidya,  V. D. and Mahler,  D. H. and Helt,  L. G. and Zatti,  L. and Liscidini,  M. and Morrison,  B. and Vernon,  Z.},
  year = {2021},
  month = apr 
}

@article{menotti_PhysRevLett.122.013904,
  title = {Nonlinear Coupling of Linearly Uncoupled Resonators},
  author = {Menotti, M. and Morrison, B. and Tan, K. and Vernon, Z. and Sipe, J. E. and Liscidini, M.},
  journal = {Phys. Rev. Lett.},
  volume = {122},
  issue = {1},
  pages = {013904},
  numpages = {5},
  year = {2019},
  month = {Jan},
  publisher = {American Physical Society},
  doi = {10.1103/PhysRevLett.122.013904},
  url = {https://link.aps.org/doi/10.1103/PhysRevLett.122.013904}
}

@article{Quesada:22,
author = {N. Quesada and L. G. Helt and M. Menotti and M. Liscidini and J. E. Sipe},
journal = {Adv. Opt. Photon.},
number = {3},
pages = {291--403},
publisher = {Optica Publishing Group},
title = {Beyond photon pairs---nonlinear quantum photonics in the high-gain regime: a tutorial},
volume = {14},
month = {Sep},
year = {2022},
url = {https://opg.optica.org/aop/abstract.cfm?URI=aop-14-3-291},
doi = {10.1364/AOP.445496},
}

@article{Su_PhysRevA.100.052301,
  title = {Conversion of Gaussian states to non-Gaussian states using photon-number-resolving detectors},
  author = {Su, Daiqin and Myers, Casey R. and Sabapathy, Krishna Kumar},
  journal = {Phys. Rev. A},
  volume = {100},
  issue = {5},
  pages = {052301},
  numpages = {32},
  year = {2019},
  month = {Nov},
  publisher = {American Physical Society},
  doi = {10.1103/PhysRevA.100.052301},
  url = {https://link.aps.org/doi/10.1103/PhysRevA.100.052301}
}

@article{seifoory_PhysRevA.105.033524,
  title = {Degenerate squeezing in a dual-pumped integrated microresonator: Parasitic processes and their suppression},
  author = {Seifoory, H. and Vernon, Z. and Mahler, D. H. and Menotti, M. and Zhang, Y. and Sipe, J. E.},
  journal = {Phys. Rev. A},
  volume = {105},
  issue = {3},
  pages = {033524},
  numpages = {14},
  year = {2022},
  month = {Mar},
  publisher = {American Physical Society},
  doi = {10.1103/PhysRevA.105.033524},
  url = {https://link.aps.org/doi/10.1103/PhysRevA.105.033524}
}

@article{Knill2001,
  title = {A scheme for efficient quantum computation with linear optics},
  volume = {409},
  ISSN = {1476-4687},
  url = {http://dx.doi.org/10.1038/35051009},
  DOI = {10.1038/35051009},
  number = {6816},
  journal = {Nature},
  publisher = {Springer Science and Business Media LLC},
  author = {Knill,  E. and Laflamme,  R. and Milburn,  G. J.},
  year = {2001},
  month = jan,
  pages = {46–52}
}

@article{PhysRevLett.119.170501,
  title = {Gaussian Boson Sampling},
  author = {Hamilton, Craig S. and Kruse, Regina and Sansoni, Linda and Barkhofen, Sonja and Silberhorn, Christine and Jex, Igor},
  journal = {Phys. Rev. Lett.},
  volume = {119},
  issue = {17},
  pages = {170501},
  numpages = {5},
  year = {2017},
  month = {Oct},
  publisher = {American Physical Society},
  doi = {10.1103/PhysRevLett.119.170501},
  url = {https://link.aps.org/doi/10.1103/PhysRevLett.119.170501}
}

@article{Aitchison1997,
  author       = {Aitchison, J.~S. and Hutchings, D.~C. and Kang, J.~U. and Stegeman, G.~I. and Villeneuve, A.},
  title        = {The nonlinear optical properties of AlGaAs at the half band gap},
  journal      = {IEEE Journal of Quantum Electronics},
  volume       = {33},
  number       = {3},
  pages        = {341--348},
  month        = mar,
  year         = {1997},
  doi          = {10.1109/3.556002},
}

@article{PhysRevA.111.023705,
  title = {Photon-pair generation via down-conversion in III-V semiconductor microrings: Modal dispersion and quasi-phase-matching},
  author = {Fontaine, Samuel E. and Vendromin, Colin and Steiner, Trevor J. and Atrli, Amirali and Thiel, Lillian and Castro, Joshua and Moody, Galan and Bowers, John and Liscidini, Marco and Sipe, J. E.},
  journal = {Phys. Rev. A},
  volume = {111},
  issue = {2},
  pages = {023705},
  numpages = {16},
  year = {2025},
  month = {Feb},
  publisher = {American Physical Society},
  doi = {10.1103/PhysRevA.111.023705},
  url = {https://link.aps.org/doi/10.1103/PhysRevA.111.023705}
}

@article{crescimanna_PhysRevA.109.023717,
  title = {Seeding Gaussian boson samplers with single photons for enhanced state generation},
  author = {Crescimanna, Valerio and Goldberg, Aaron Z. and Heshami, Khabat},
  journal = {Phys. Rev. A},
  volume = {109},
  issue = {2},
  pages = {023717},
  numpages = {14},
  year = {2024},
  month = {Feb},
  publisher = {American Physical Society},
  doi = {10.1103/PhysRevA.109.023717},
  url = {https://link.aps.org/doi/10.1103/PhysRevA.109.023717}
}

@article{drago_PhysRevA.110.023710,
  title = {Deconstructing squeezed light: Schmidt decomposition versus the Whittaker-Shannon interpolation},
  author = {Drago, C. and Sipe, J. E.},
  journal = {Phys. Rev. A},
  volume = {110},
  issue = {2},
  pages = {023710},
  numpages = {31},
  year = {2024},
  month = {Aug},
  publisher = {American Physical Society},
  doi = {10.1103/PhysRevA.110.023710},
  url = {https://link.aps.org/doi/10.1103/PhysRevA.110.023710}
}

@article{Zielnicki2018,
  title = {Joint spectral characterization of photon-pair sources},
  volume = {65},
  ISSN = {1362-3044},
  url = {http://dx.doi.org/10.1080/09500340.2018.1437228},
  DOI = {10.1080/09500340.2018.1437228},
  number = {10},
  journal = {Journal of Modern Optics},
  publisher = {Informa UK Limited},
  author = {Zielnicki,  Kevin and Garay-Palmett,  Karina and Cruz-Delgado,  Daniel and Cruz-Ramirez,  Hector and O’Boyle,  Michael F. and Fang,  Bin and Lorenz,  Virginia O. and U’Ren,  Alfred B. and Kwiat,  Paul G.},
  year = {2018},
  month = feb,
  pages = {1141–1160}
}

@misc{bulmer2025simulatinglossypartiallydistinguishable,
      title={Simulating lossy and partially distinguishable quantum optical circuits: theory, algorithms and applications to experiment validation and state preparation}, 
      author={Jacob F. F. Bulmer and Javier Martínez-Cifuentes and Bryn A. Bell and Nicolás Quesada},
      year={2025},
      eprint={2412.17742},
      archivePrefix={arXiv},
      primaryClass={quant-ph},
      url={https://arxiv.org/abs/2412.17742}, 
}

@article{Vernon:17,
author = {Z. Vernon and M. Menotti and C. C. Tison and J. A. Steidle and M. L. Fanto and P. M. Thomas and S. F. Preble and A. M. Smith and P. M. Alsing and M. Liscidini and J. E. Sipe},
journal = {Opt. Lett.},
number = {18},
pages = {3638--3641},
publisher = {Optica Publishing Group},
title = {Truly unentangled photon pairs without spectral filtering},
volume = {42},
month = {Sep},
year = {2017},
url = {https://opg.optica.org/ol/abstract.cfm?URI=ol-42-18-3638},
doi = {10.1364/OL.42.003638},
}

@article{Helt:10_SFWMringresonator,
author = {L. G. Helt and Zhenshan Yang and Marco Liscidini and J. E. Sipe},
journal = {Opt. Lett.},
number = {18},
pages = {3006--3008},
publisher = {Optica Publishing Group},
title = {Spontaneous four-wave mixing in microring resonators},
volume = {35},
month = {Sep},
year = {2010},
url = {https://opg.optica.org/ol/abstract.cfm?URI=ol-35-18-3006},
doi = {10.1364/OL.35.003006},
}

@article{leger_PhysRevResearch.5.023131,
  title = {Amplification of cascaded down-conversion by reusing photons with a switchable cavity},
  author = {Leger, Alexandre Z. and Gambhir, Samridhi and L\'eg\`ere, Julien and Hamel, Deny R.},
  journal = {Phys. Rev. Res.},
  volume = {5},
  issue = {2},
  pages = {023131},
  numpages = {6},
  year = {2023},
  month = {May},
  publisher = {American Physical Society},
  doi = {10.1103/PhysRevResearch.5.023131},
  url = {https://link.aps.org/doi/10.1103/PhysRevResearch.5.023131}
}

@article{banicTOPDC_PhysRevA.106.013710,
  title = {Resonant and nonresonant integrated third-order parametric down-conversion},
  author = {Banic, Milica and Liscidini, Marco and Sipe, J. E.},
  journal = {Phys. Rev. A},
  volume = {106},
  issue = {1},
  pages = {013710},
  numpages = {19},
  year = {2022},
  month = {Jul},
  publisher = {American Physical Society},
  doi = {10.1103/PhysRevA.106.013710},
  url = {https://link.aps.org/doi/10.1103/PhysRevA.106.013710}
}

@article{Moebius:16,
author = {Michael G. Moebius and Felipe Herrera and Sarah Griesse-Nascimento and Orad Reshef and Christopher C. Evans and Gian Giacomo Guerreschi and Al\'{a}n Aspuru-Guzik and Eric Mazur},
journal = {Opt. Express},
number = {9},
pages = {9932--9954},
publisher = {Optica Publishing Group},
title = {Efficient photon triplet generation in integrated nanophotonic waveguides},
volume = {24},
month = {May},
year = {2016},
url = {https://opg.optica.org/oe/abstract.cfm?URI=oe-24-9-9932},
doi = {10.1364/OE.24.009932},
}

@article{Lloyd_PhysRevLett.82.1784,
  title = {Quantum Computation over Continuous Variables},
  author = {Lloyd, Seth and Braunstein, Samuel L.},
  journal = {Phys. Rev. Lett.},
  volume = {82},
  issue = {8},
  pages = {1784--1787},
  numpages = {0},
  year = {1999},
  month = {Feb},
  publisher = {American Physical Society},
  doi = {10.1103/PhysRevLett.82.1784},
  url = {https://link.aps.org/doi/10.1103/PhysRevLett.82.1784}
}

@article{Baboux:23,
author = {F. Baboux and G. Moody and S. Ducci},
journal = {Optica},
number = {7},
pages = {917--931},
publisher = {Optica Publishing Group},
title = {Nonlinear integrated quantum photonics with AlGaAs},
volume = {10},
month = {Jul},
year = {2023},
url = {https://opg.optica.org/optica/abstract.cfm?URI=optica-10-7-917},
doi = {10.1364/OPTICA.481385},
}

@article{bethebreit_PhysRev.93.888,
  title = {Ingoing Waves in Final State of Scattering Problems},
  author = {Breit, G. and Bethe, H. A.},
  journal = {Phys. Rev.},
  volume = {93},
  issue = {4},
  pages = {888--890},
  numpages = {0},
  year = {1954},
  month = {Feb},
  publisher = {American Physical Society},
  doi = {10.1103/PhysRev.93.888},
  url = {https://link.aps.org/doi/10.1103/PhysRev.93.888}
}

@article{steiner_PRXQuantum.2.010337,
  title = {Ultrabright Entangled-Photon-Pair Generation from an $\mathrm{Al}\mathrm{Ga}\mathrm{As}$-On-Insulator Microring Resonator},
  author = {Steiner, Trevor J. and Castro, Joshua E. and Chang, Lin and Dang, Quynh and Xie, Weiqiang and Norman, Justin and Bowers, John E. and Moody, Galan},
  journal = {PRX Quantum},
  volume = {2},
  issue = {1},
  pages = {010337},
  numpages = {11},
  year = {2021},
  month = {Mar},
  publisher = {American Physical Society},
  doi = {10.1103/PRXQuantum.2.010337},
  url = {https://link.aps.org/doi/10.1103/PRXQuantum.2.010337}
}

@article{Castro2022,
  author       = {Castro, Joshua E. and Steiner, Trevor J. and Thiel, Lillian and Dinkelacker, Alex and McDonald, Corey and Pintus, Paolo and Chang, Lin and Bowers, John E. and Moody, Galan},
  title        = {Expanding the Quantum Photonic Toolbox in AlGaAsOI},
  journal      = {APL Photonics},
  volume       = {7},
  number       = {9},
  pages        = {096103},
  year         = {2022},
  publisher    = {AIP Publishing},
  doi          = {10.1063/5.0098984},
  url          = {https://doi.org/10.1063/5.0098984}
}

@article{pu2022design,
  title={Design and fabrication of AlGaAs-on-insulator microring resonators for nonlinear photonics},
  author={Pu, Minhao and Hakkarainen, Teemu and Kivshar, Yuri and Rottwitt, Karsten},
  journal={IEEE Journal of Selected Topics in Quantum Electronics},
  volume={28},
  number={5},
  pages={1--9},
  year={2022},
  publisher={IEEE},
  doi={10.1109/JSTQE.2022.3159792}
}

@article{PhysRevA.106.043707,
  title = {Two strategies for modeling nonlinear optics in lossy integrated photonic structures},
  author = {Banic, Milica and Zatti, Luca and Liscidini, Marco and Sipe, J. E.},
  journal = {Phys. Rev. A},
  volume = {106},
  issue = {4},
  pages = {043707},
  numpages = {18},
  year = {2022},
  month = {Oct},
  publisher = {American Physical Society},
  doi = {10.1103/PhysRevA.106.043707},
  url = {https://link.aps.org/doi/10.1103/PhysRevA.106.043707}
}

@ARTICLE{tuneableQ_10930539,
  author={Pagano, Paula L. and Borghi, Massimo and Moroni, Federica and Viola, Alice and Malaspina, Francesco and Liscidini, Marco and Bajoni, Daniele and Galli, Matteo},
  journal={Journal of Lightwave Technology}, 
  title={Selective Linewidth Control in a Micro-Resonator With a Resonant Interferometric Coupler}, 
  year={2025},
  volume={43},
  number={12},
  pages={5731-5737},
  doi={10.1109/JLT.2025.3552182}}

@article{Hamel2014,
  title = {Direct generation of three-photon polarization entanglement},
  volume = {8},
  ISSN = {1749-4893},
  url = {http://dx.doi.org/10.1038/nphoton.2014.218},
  DOI = {10.1038/nphoton.2014.218},
  number = {10},
  journal = {Nature Photonics},
  publisher = {Springer Science and Business Media LLC},
  author = {Hamel,  Deny R. and Shalm,  Lynden K. and H\"{u}bel,  Hannes and Miller,  Aaron J. and Marsili,  Francesco and Verma,  Varun B. and Mirin,  Richard P. and Nam,  Sae Woo and Resch,  Kevin J. and Jennewein,  Thomas},
  year = {2014},
  month = sep,
  pages = {801–807}
}

@article{Helt:12,
author = {Lukas G. Helt and Marco Liscidini and John E. Sipe},
journal = {J. Opt. Soc. Am. B},
number = {8},
pages = {2199--2212},
publisher = {Optica Publishing Group},
title = {How does it scale? Comparing quantum and classical nonlinear optical processes in integrated devices},
volume = {29},
month = {Aug},
year = {2012},
url = {https://opg.optica.org/josab/abstract.cfm?URI=josab-29-8-2199},
doi = {10.1364/JOSAB.29.002199},
}

@article{Yanagimoto:22_mesoscopic_nG,
author = {Ryotatsu Yanagimoto and Edwin Ng and Atsushi Yamamura and Tatsuhiro Onodera and Logan G. Wright and Marc Jankowski and M. M. Fejer and Peter L. McMahon and Hideo Mabuchi},
journal = {Optica},
number = {4},
pages = {379--390},
publisher = {Optica Publishing Group},
title = {Onset of non-Gaussian quantum physics in pulsed squeezing with mesoscopic fields},
volume = {9},
month = {Apr},
year = {2022},
url = {https://opg.optica.org/optica/abstract.cfm?URI=optica-9-4-379},
doi = {10.1364/OPTICA.447782},
}

@article{PhysRevApplied.23.014003,
  title = {Direct generation of multiphoton hyperentanglement},
  author = {Zhao, Peng and Ying, Jia-Wei and Yang, Meng-Ying and Zhong, Wei and Du, Ming-Ming and Shen, Shu-Ting and Li, Yun-Xi and Zhang, An-Lei and Zhou, Lan and Sheng, Yu-Bo},
  journal = {Phys. Rev. Appl.},
  volume = {23},
  issue = {1},
  pages = {014003},
  numpages = {20},
  year = {2025},
  month = {Jan},
  publisher = {American Physical Society},
  doi = {10.1103/PhysRevApplied.23.014003},
  url = {https://link.aps.org/doi/10.1103/PhysRevApplied.23.014003}
}

@article{Hbel2010_nature,
  title = {Direct generation of photon triplets using cascaded photon-pair sources},
  volume = {466},
  ISSN = {1476-4687},
  url = {http://dx.doi.org/10.1038/nature09175},
  DOI = {10.1038/nature09175},
  number = {7306},
  journal = {Nature},
  publisher = {Springer Science and Business Media LLC},
  author = {H\"{u}bel,  Hannes and Hamel,  Deny R. and Fedrizzi,  Alessandro and Ramelow,  Sven and Resch,  Kevin J. and Jennewein,  Thomas},
  year = {2010},
  month = jul,
  pages = {601–603}
}

@article{Krapick:16,
author = {Stephan Krapick and Benjamin Brecht and Harald Herrmann and Viktor Quiring and Christine Silberhorn},
journal = {Opt. Express},
number = {3},
pages = {2836--2849},
publisher = {Optica Publishing Group},
title = {On-chip generation of photon-triplet states},
volume = {24},
month = {Feb},
year = {2016},
url = {https://opg.optica.org/oe/abstract.cfm?URI=oe-24-3-2836},
doi = {10.1364/OE.24.002836},
}

@article{bib:Adachi1985,
    author = {Adachi, S.},
    title = "{GaAs, AlAs, and AlxGa1-xAs: Material parameters for use in research and device applications}",
    journal = {Journal of Applied Physics},
    volume = {58},
    number = {3},
    pages = {R1-R29},
    year = {1985},
    month = {08},
    issn = {0021-8979},
    url = {https://doi.org/10.1063/1.336070},
}

@mastersthesis{Johnson2019FullVectorialAlGaAs,
  author       = {Johnson, Kyle Anthony},
  title        = {A Full-Vectorial Theory of Third-Order Nonlinear Optical Effects in Aluminum Gallium Arsenide Waveguides},
  school       = {University of Toronto},
  year         = {2019},
  address      = {Toronto, Ontario, Canada},
  type         = {M.A.Sc.\ Thesis},
  url          = {https://tspace.library.utoronto.ca/handle/1807/100634},
}

@article{AfsharV.:09,
author = {Shahraam Afshar V. and Tanya M. Monro},
journal = {Opt. Express},
number = {4},
pages = {2298--2318},
publisher = {Optica Publishing Group},
title = {A full vectorial model for pulse propagation in emerging waveguides with subwavelength structures part I: Kerr nonlinearity},
volume = {17},
month = {Feb},
year = {2009},
url = {https://opg.optica.org/oe/abstract.cfm?URI=oe-17-4-2298},
doi = {10.1364/OE.17.002298},
}

@article{DominguezSerna2020,
  author = {Francisco A. Dom{\'\i}nguez-Serna and Alfred B. U'Ren and Karina Garay-Palmett},
  title = {Third-order parametric down-conversion: A stimulated approach},
  journal = {Phys. Rev. A},
  volume = {101},
  issue = {3},
  pages = {033813},
  year = {2020},
  doi = {10.1103/PhysRevA.101.033813},
  url = {https://doi.org/10.1103/PhysRevA.101.033813}
}

@book{yanagimoto2023quantum,
  title={Quantum Dynamics of Broadband Nonlinear Photonics: From Phenomenology to Function},
  author={Yanagimoto, R. and Mabuchi, H. and Hayden, P.M. and Safavi-Naeini, A.H. and Stanford University. School of Humanities and Sciences and Stanford University. Department of Applied Physics},
  url={https://books.google.ca/books?id=ejvRzwEACAAJ},
  year={2023},
  publisher={Stanford University}
}

@article{Shukhin:24,
author = {Anatoly Shukhin and Inbar Hurvitz and Sivan Trajtenberg-Mills and Ady Arie and Hagai Eisenberg},
journal = {Opt. Express},
number = {6},
pages = {10158--10174},
publisher = {Optica Publishing Group},
title = {Two-dimensional control of a biphoton joint spectrum},
volume = {32},
month = {Mar},
year = {2024},
url = {https://opg.optica.org/oe/abstract.cfm?URI=oe-32-6-10158},
doi = {10.1364/OE.510497},
}

@article{May:19,
author = {Stuart May and Michael Kues and Matteo Clerici and Marc Sorel},
journal = {Opt. Lett.},
number = {6},
pages = {1339--1342},
publisher = {Optica Publishing Group},
title = {Second-harmonic generation in AlGaAs-on-insulator waveguides},
volume = {44},
month = {Mar},
year = {2019},
url = {https://opg.optica.org/ol/abstract.cfm?URI=ol-44-6-1339},
doi = {10.1364/OL.44.001339},
}

@article{10.1063/1.3119629,
    author = {Wagner, Sean J. and Holmes, Barry M. and Younis, Usman and Helmy, Amr S. and Aitchison, J. Stewart and Hutchings, David C.},
    title = {Continuous wave second-harmonic generation using domain-disordered quasi-phase matching waveguides},
    journal = {Applied Physics Letters},
    volume = {94},
    number = {15},
    pages = {151107},
    year = {2009},
    month = {04},
    issn = {0003-6951},
    doi = {10.1063/1.3119629},
    url = {https://doi.org/10.1063/1.3119629},
}

@article{Hutchings:97,
author = {D. C. Hutchings and J. S. Aitchison and J. M. Arnold},
journal = {J. Opt. Soc. Am. B},
number = {4},
pages = {869--879},
publisher = {Optica Publishing Group},
title = {Nonlinear refractive coupling and vector solitons in anisotropic cubic media},
volume = {14},
month = {Apr},
year = {1997},
url = {https://opg.optica.org/josab/abstract.cfm?URI=josab-14-4-869},
doi = {10.1364/JOSAB.14.000869},
}

@article{Hutchings:95,
author = {D. C. Hutchings and J. S. Aitchison and B. S. Wherrett and G. T. Kennedy and W. Sibbett},
journal = {Opt. Lett.},
number = {9},
pages = {991--993},
publisher = {Optica Publishing Group},
title = {Polarization dependence of ultrafast nonlinear refraction in an AlGaAs waveguide at the half-band gap},
volume = {20},
month = {May},
year = {1995},
url = {https://opg.optica.org/ol/abstract.cfm?URI=ol-20-9-991},
doi = {10.1364/OL.20.000991},
}

@article{PRXQuantum.6.010338,
  title = {Versatile Chip-Scale Platform for High-Rate Entanglement Generation Using an $\mathrm{Al}\mathrm{Ga}\mathrm{As}$ Microresonator Array},
  author = {Pang, Yiming and Castro, Joshua E. and Steiner, Trevor J. and Duan, Liao and Tagliavacche, Noemi and Borghi, Massimo and Thiel, Lillian and Lewis, Nicholas and Bowers, John E. and Liscidini, Marco and Moody, Galan},
  journal = {PRX Quantum},
  volume = {6},
  issue = {1},
  pages = {010338},
  numpages = {15},
  year = {2025},
  month = {Mar},
  publisher = {American Physical Society},
  doi = {10.1103/PRXQuantum.6.010338},
  url = {https://link.aps.org/doi/10.1103/PRXQuantum.6.010338}
}

@misc{Vendromin25,
      title={Non-Gaussian states via pump-depleted SPDC}, 
      author={Colin Vendromin and Samuel E. Fontaine and J. E. Sipe},
      year={2025},
      eprint={2510.06498},
      archivePrefix={arXiv},
      primaryClass={quant-ph},
      url={https://arxiv.org/abs/2510.06498}, 
}

\pagebreak
\twocolumngrid
\appendix

\section{\label{appendix:asyfields}Linear physics}
The linear Hamiltonian for the rings, point-coupled to a waveguide as shown in Fig. \ref{fig:rings_modes}a, consists of the sum of the isolated ring Hamiltonians, three waveguide Hamiltonians, two coupling Hamiltonians to the bus waveguide, and two coupling Hamiltonians to the phantom waveguides ($\lambda\in\{\mathrm{ac,ph1,ph2}\}$, $\ell\in\{1,2\}$):
\begin{equation}
H_{\mathrm{L}} = \sum_{\lambda}H^{\mathrm{wg}}_{\lambda} + \sum_\ell \left( H^{\mathrm{ring}}_{(\ell)} + H^{\mathrm{cpl}}_{\mathrm{ac}(\ell)} + H^{\mathrm{cpl}}_{\mathrm{ph}(\ell)} \right).
    \label{eq:Linear}
\end{equation}
The $\ell^{\mathrm{th}}$ isolated ring Hamiltonian is
\begin{equation}
\begin{split}
    &H^{(\ell)}_{\mathrm{ring}} = \sum_{u} \hbar \omega_u^{(\ell)} b^{(\ell)\dagger}_u b_u^{(\ell)},
\end{split}
    \label{eq:RingHamiltonian}
\end{equation}
with the $\ell^{\mathrm{th}}$ ring's resonant frequencies denoted by $\omega_u^{(\ell)}$, with the usual commutation relations
\begin{equation}
\left[b_u^{(\ell)},b_{u'}^{(\ell')\dagger}\right] = \delta_{uu'} \delta_{\ell\ell'}.
\label{eq:CommuatationRelations_bJ}
\end{equation}
The index $u$ denotes a frequency range with center resonance $\omega_u$ where up to single a resonance of each ring can be found; e.g., in the frequency range $S_{1}$, the first ring could have a resonance with resonant frequency $\omega_{S_{1}}^{(1)}$ or not, and similarly for $\omega_{S_{1}}^{(2)}$, and in general both frequencies can be different. 

The Hamiltonian for the isolated $\lambda$ waveguide is
\begin{equation}
\begin{split}
    &H_{\lambda}^{\mathrm{wg}}  = \sum_{u} \Bigg[\int\hbar \omega_u \psi^{\lambda\dagger}_u(s) \psi^\lambda_u(s)\mathrm{d}s \\
    & -\frac{i\hbar v_{u}}{2} \int \left(\psi^{\lambda\dagger}_u(s) \frac{\partial\psi^\lambda_u(s)}{\partial s} - \frac{\partial\psi_u^{\lambda\dagger}(s)}{\partial s}\psi^\lambda_u(s) \right) \mathrm{d}s\Bigg], 
    \end{split}
\end{equation}
where 
\begin{equation}
    \psi^\lambda_u(s) = \int \frac{\mathrm{d}k}{\sqrt{2\pi}} a^{\lambda}_u(k)e^{i(k-K_u)s},
    \label{eq:psi(s)}
\end{equation}
is the channel operator associated with the resonant frequency range of the resonator \cite{Quesada:22}, and has the usual commutation relations
\begin{equation}
\left[a_u^\lambda(k),\left(a_{u'}^{\lambda'}(k)\right)^\dagger\right] = \delta_{uu'}\delta_{\lambda\lambda'}\delta(k-k').
    \label{eq:a(k)Commutator}
\end{equation}
We introduced $v_u = \left(\partial\omega_{uk}/\partial k\right)_{K_u}$ and $K_u$, which are the group velocity and a reference wavenumber in the waveguides associated with the frequency range $u$, over which group velocity dispersion is neglected. For convenience, $K_u$ is often chosen as a resonant wavenumber of a ring $\kappa_u^{(\ell)}=2\pi m_u^{(\ell)} / \mathcal{L}^{(\ell)}$, where $m_u^{(\ell)}$ and $\mathcal{L}^{(\ell)}$ are the resonant mode numbers and circumference of the $\ell^{\mathrm{th}}$ ring, respectively. The waveguides are assumed to be identical, and so there is no waveguide dependence on $v_u$ and $K_u$, but this can
easily be generalized. We write the dispersion relation around the frequency $\omega_u$ of the frequency range of interest
\begin{equation}
\omega_{uk} = \omega_{u} + v_u(k-K_u).
    \label{eq:DispRel}
\end{equation}
Using $s$ to denote the distance along each waveguide (bus waveguide and phantom waveguides), for the bus waveguide we take the coupling point to the first (second) ring to be at $s=-a$ ($s=+a$), and for both phantom waveguides the coupling point is at $s=0$. The coupling Hamiltonians for the $\ell^{\mathrm{th}}$ ring are
\begin{equation}
\begin{split}
    & H_{\mathrm{ac}(\ell)}^{\mathrm{cpl}} = \sum_{u} \left[\hbar \left(\gamma^{\mathrm{ac}(\ell)}_u\right)^* b_u^{(\ell)\dagger} \psi_u(\pm a) + \mathrm{H.c.}\right],\\
    & H_{\mathrm{ph}(\ell)}^{\mathrm{cpl}} = \sum_{u} \left[\hbar \left(\gamma^{\mathrm{ph}(\ell)}_u\right)^* b_u^{(\ell)\dagger} {\varphi^{(\ell)}_u} (0) + \mathrm{H.c.}\right],\\
\end{split}
    \label{eq:CplHamiltonians}
\end{equation}
where we take $-a$ ($+a$) for $\ell=1$ ($\ell=2$). We define decay rates to the waveguides, quality factors, and escape efficiencies from the coupling coefficients $\gamma_u^{\lambda(\ell)}$ from the above Hamiltonians \cite{Quesada:22}: 
\begin{align}
\Gamma_u^{\lambda(\ell)} & \equiv \frac{|\gamma_u^{\lambda(\ell)}|^2}{2v_u} = \frac{\omega_u^{(\ell)}}
     {2Q_u^{\lambda(\ell)}},\\
\bar{\Gamma}^{(\ell)}_u &\equiv \Gamma_u^{\mathrm{ac}(\ell)} + \Gamma_u^{\mathrm{ph}(\ell)} = \frac{\omega_u^{(\ell)}}{2Q_u^{\mathrm{(load)}{(\ell)}}},\label{eq:GammaDef}\\
\eta_u^{\lambda(\ell)}&\equiv\frac{\Gamma_u^{\lambda(\ell)} }{ \bar{\Gamma}_u^{(\ell)}} = \frac{Q_u^{\mathrm{(load)}{(\ell)}}}{Q_u^{\lambda(\ell)}}.
\label{eq:EscEff}
\end{align}

{\section{\label{appendix:NL}Nonlinear physics}}
The relevant Hamiltonians for the cascaded SFWM processes are
\begin{align}
    H_{\text{SFWM1}} &= -\frac{1}{4\epsilon_0} \frac{4!}{2!1!1!}\int \mathrm{d}\boldsymbol{r} \Gamma^{(3)}_{ijkl}(\boldsymbol{r}) 
    D^{\text{out},i}_{S_{1}}(\boldsymbol{r}) D^{\text{out},j}_{I}(\boldsymbol{r}) 
    \nonumber \\ & \hspace{3cm} \times D^{\text{in},k}_{P_{1}}(\boldsymbol{r}) D^{\text{in},l}_{P_{1}}(\boldsymbol{r})
    ,\label{eq:HSFWM1_def}\\
    H_{\text{SFWM2}} &= -\frac{1}{4\epsilon_0} \frac{4!}{1!1!1!1!}\int \mathrm{d}\boldsymbol{r} \Gamma^{(3)}_{ijkl}(\boldsymbol{r}) 
    D^{\text{out},i}_{S_{2}}(\boldsymbol{r}) D^{\text{out},j}_{S_{3}}(\boldsymbol{r}) 
    \nonumber \\ & \hspace{3cm} \times D^{\text{out},k}_{I}(\boldsymbol{r}) D^{\text{in},l}_{P_{2}}(\boldsymbol{r}).
    \label{eq:HSFWM2_def}
\end{align}
We note that in deriving Eqs. \eqref{eq:HSFWM1_def}-\eqref{eq:HSFWM2_def} full permutation symmetry of the $\Gamma^{(3)}$ tensor is assumed \cite{Quesada:22}, and in Appendix \ref{Supp} we briefly discuss the effects of this assumption on our results.

We move to the interaction picture using 
$H_{\mathrm{NL}}^{(I)}(t) = e^{iH_{\mathrm{L}}t/\hbar} H_{\mathrm{NL}} e^{-iH_{\mathrm{L}}t/\hbar}$, where $H_{\mathrm{L}}$ is the linear Hamiltonian. In terms of the asymptotic field operators, we can write the linear Hamiltonian as
\begin{align}
    H_{\mathrm{L}} = \sum_{u,\lambda} \int \mathrm{d}k \hbar \omega_{uk} \left(a_{u}^{\text{in/out},\lambda}(k)\right)^{\dagger} a^{\text{in/out},\lambda}_{u}(k),
\end{align}
so the time dependence of the ladder operators in the interaction picture is trivial; we have 
\begin{align}
    \boldsymbol{D}_u^{\text{in/out}(I)}(\boldsymbol{r},t) = &\sum_{\lambda} \int \mathrm{d}k \boldsymbol{\mathcal{D}}_{uk}^{\text{in/out}, \lambda}(\boldsymbol{r}) \nonumber \\
    & \times a_u^{\text{in/out},\lambda}(k) e^{-i\omega_{uk}t} + \mathrm{H.c..} 
    \label{eq:Dasy_I_def}
\end{align}
We treat the pump fields $P_{1}$ and $P_{2}$ classically, taking $a_{u}^{\mathrm{in,ac}}(k) \rightarrow \alpha_{u} \phi_{u}(k)$, where $\phi_{u}(k)$ is a normalized function that defines the pump spectrum, and $\alpha_{u}$ is a constant set to yield the correct pulse energy. In this work, we only have pump fields entering through one channel (the bus waveguide), so the asymptotic-in fields for the pump fields have the form
\begin{align}
    \boldsymbol{D}_u^{\text{in}(I)}(\boldsymbol{r},t) = \alpha_u \int \mathrm{d}k \boldsymbol{\mathcal{D}}_{uk}^{\text{in}, \lambda}(\boldsymbol{r}) \phi_u(k) e^{-i\omega_{uk}t} + \mathrm{H.c..} \label{eq:Dasy_pump_def}
\end{align}
With the fields in the interaction picture, and recalling Eq. \eqref{eq:HSFWM1_def}, we find
\begin{align}
    H_{\text{SFWM1}}^{(I)}&(t) = \frac{-\alpha_{P_{1}}^2}{3\epsilon_0} \int \mathrm{d} \boldsymbol{r} \Gamma_{ijkl}^{(3)}(\boldsymbol{r})\nonumber\\
    \times &\left(  \sum_{\lambda} \int \mathrm{d}k {\mathcal{D}}_{S_{1} k}^{\text{out}, \lambda,i}(\boldsymbol{r}) a_{S_{1}}^{\text{out},\lambda}(k)e^{-i\omega_{S_{1} k}t} + \mathrm{H.c.} \right)\nonumber\\
     \times &\left(  \sum_{\lambda'} \int \mathrm{d}k {\mathcal{D}}_{I k}^{\text{out}, \lambda',j}(\boldsymbol{r}) a_{I}^{\text{out},\lambda'}(k)e^{-i\omega_{I k}t} + \mathrm{H.c.} \right) \nonumber\\
     \times &\left(  \int \mathrm{d}k {\mathcal{D}}_{P_{1} k}^{\text{in},k}(\boldsymbol{r}) \phi_{P_{1}}^{\text{in}}(k)e^{-i\omega_{P_{1} k}t} + \mathrm{H.c.} \right) \nonumber \\
    \times &\left( \int \mathrm{d}k {\mathcal{D}}_{P_{1} k}^{\text{in}, l}(\boldsymbol{r}) \phi_{P_{1}}^{\text{in}}(k)e^{-i\omega_{P_{1} k}t} + \mathrm{H.c.} \right).
    \end{align}
Rearranging and neglecting the fast-rotating terms, we have
\begin{align}
    H^{(I)}_{\text{SFWM1}}(t)& = -\sum_{\lambda\lambda'} \int \mathrm{d}k_1 \mathrm{d}k_2 \mathrm{d}k_3 \mathrm{d}k_4  K^{\lambda\lambda'}_{1}(k_1,k_2,k_3,k_4) \nonumber\\ 
    &\times e^{-i\Omega_{1}(k_1,k_2,k_3,k_4)t}  \alpha^2_{P_{1}}\phi_{P_{1}}(k_3) \phi_{P_{1}}(k_4)\nonumber\\
    & \times \left[ a^{\text{out}, \lambda}_{S_{1}}(k_1) a^{\text{out}, \lambda'}_{I}(k_2) \right]^{\dagger}, 
    \label{eq:HSFWM1} 
\end{align}
and similarly, using Eq. \eqref{eq:HSFWM2_def}, for the second SFWM interaction, we obtain
\begin{align}
    H^{(I)}_{\text{SFWM2}}(t)& = -\sum_{\mu\mu'\mu''} \int \mathrm{d}k_1 \mathrm{d}k_2 \mathrm{d}k_3 \mathrm{d}k_4  K^{\mu \mu' \mu''}_{2}(k_1,k_2,k_3,k_4)\nonumber\\ 
    &\times e^{-i\Omega_{2}(k_1,k_2,k_3,k_4)t} \alpha_{P_{2}} \phi_{P_{2}}(k_4) \nonumber\\
    &\times a^{\text{out}, \mu}_{I}(k_3) \left[ a^{\text{out}, \mu'}_{S_{2}}(k_1) a^{\text{out}, \mu''}_{S_{3}}(k_2) \right]^{\dagger}.  \label{eq:HSFWM2}
\end{align} 
We have introduced  
\begin{align}
    \Omega_1(k_1,...,k_4) &= \omega_{P_{1} k_3} + \omega_{P_{1} k_4} - \omega_{S_{1} k_1} - \omega_{I k_2}\\
    \Omega_2(k_1,...,k_4) &= \omega_{I k_3} + \omega_{P_{2} k_4} - \omega_{S_{2} k_1} - \omega_{S_{3} k_2}
\end{align}
and
\begin{align}
        K_1^{\lambda\lambda'}(k_1,...,k_4) = \frac{3}{\epsilon_0} \int \mathrm{d}\boldsymbol{r} & \Gamma_{ijkl}^{(3)}  \left[ \mathcal{D}^{\text{out},\lambda,i}_{S_{1} k_1}\mathcal{D}^{\text{out},\lambda',j}_{I k_2}\right]^{*}\nonumber \\ &   \times\mathcal{D}^{\text{in},k}_{P_{1} k_3}\mathcal{D}^{\text{in},l}_{P_{1} k_4}\label{eq:K1}\\ 
         K_2^{\mu\mu'\mu''}(k_1,...,k_4) = \frac{6}{\epsilon_0} \int \mathrm{d}\boldsymbol{r} &\Gamma_{ijkl}^{(3)}  \left[ \mathcal{D}^{\text{out},\mu',i}_{S_{2} k_1}\mathcal{D}^{\text{out},\mu'',j}_{S_{3} k_2}\right]^{*}\nonumber \\ 
        &  \times \mathcal{D}^{\text{out},\mu,k}_{I k_3}\mathcal{D}^{\text{in,ac},l}_{P_{2} k_4} \label{eq:K2},
\end{align}
where the position dependence of the integrands was dropped for brevity.

Focusing on the ``low-gain" regime for both SFWM processes, from Eq. \eqref{eq:SE} the generated state can be written as 
\begin{align}
    \ket{\psi(t)} &\approx \ket{\text{vac}} -\frac{i}{\hbar} \int_{t_0}^{t} \mathrm{d}t' H_{\mathrm{NL}}^{(I)}(t')\ket{\text{vac}} \nonumber \\ &- \frac{1}{\hbar^2} \int^{t}_{t_0} \mathrm{d}t' \int^{t'}_{t_0} \mathrm{d}t'' H^{(I)}_{NL}(t') H^{(I)}_{\mathrm{NL}}(t'') \ket{\text{vac}}, \nonumber\\
    &= \ket{\text{vac}} -\frac{i}{\hbar} \int_{t_0}^{t} \mathrm{d}t' H_{\text{SFWM1}}^{(I)}(t')\ket{\text{vac}} \nonumber \\ \nonumber &- \frac{1}{\hbar^2} \int^{t}_{t_0} \mathrm{d}t' \int^{t'}_{t_0} \mathrm{d}t'' H^{(I)}_{\mathrm{SFWM}2}(t') H^{(I)}_{\mathrm{SFWM}1}(t'') \ket{\text{vac}}\\
     &- \frac{1}{\hbar^2} \int^{t}_{t_0} \mathrm{d}t' \int^{t'}_{t_0} \mathrm{d}t'' H^{(I)}_{\mathrm{SFWM}1}(t') H^{(I)}_{\mathrm{SFWM}1}(t'') \ket{\text{vac}}. \label{eq:ket_O2}
 \end{align}
In the second line, we have used the fact that $H_{\mathrm{SFWM}2}\ket{\text{vac}} = 0$. 

We neglect the final term in Eq. \eqref{eq:ket_O2}, which describes the generation of two pairs in the first ring. Inserting Eqs. \eqref{eq:HSFWM1} and \eqref{eq:HSFWM2} into \eqref{eq:ket_O2} and taking $t_0\rightarrow -\infty$, we obtain Eq. \eqref{eq:ket}, with
\begin{align}
    \varphi^{\lambda \lambda'}(k_1,k_2;t) &= \frac{i \alpha_{P_{1}}^2}{\hbar \beta} \int \mathrm{d}k_3 \mathrm{d}k_4 K_1^{\lambda \lambda'}(k_1,...,k_4) \phi_{P_{1}}(k_3) \nonumber\\
    & \times \phi_{P_{1}}(k_4) \int_{-\infty}^{t} \mathrm{d}t' e^{-i\Omega_1(k_1,...,k_4)t'} \label{eq:BWF_def}
\end{align}
and
\begin{align}
\Psi^{\mu \mu' \mu''}(k'_1, k_2, k_3;t) &= -\frac{i \alpha_{P_{2}} \beta}{\hbar \sigma} \int \mathrm{d}k_1 \mathrm{d}k_4 K_2^{\mu \mu' \mu''}(k_1,...,k_4) \nonumber \\
& \times \phi_{P_{2}}(k_4) \int_{-\infty}^{t} \mathrm{d}t' \sum_{\lambda'} \varphi^{\mu \lambda'}(k'_1,k_1;t') \nonumber \\
& \times e^{-i\Omega_2(k_1,...,k_4)t'}. \label{eq:TWF_def}
\end{align}
The wavefunctions are normalized according to
\begin{align}
    &\sum_{\lambda \lambda'} \int \mathrm{d}k_1 \mathrm{d}k_2 |\varphi^{\lambda\lambda'}(k_1,k_2;t\rightarrow \infty)|^2 = 1 \label{eq:BWF_full_norm}
\end{align}
and
\begin{align}
    \hspace{-0.3cm}\sum_{\mu \mu'\mu''} \int \mathrm{d}k_1 \mathrm{d}k_2 \mathrm{d}k_3 |\Psi^{\mu\mu'\mu''}(k_1,k_2,k_3;t\rightarrow \infty)|^2 = 1. \label{eq:TWF_full_norm}
\end{align}

\onecolumngrid

{

\section{\label{Supp}Nonlinear parameters and tensors}
We can write Eqs. \eqref{eq:K1} and \eqref{eq:K2} in terms of nonlinear parameters $\gamma_{\mathrm{NL}}$ \cite{Quesada:22}. For simplicity, we write the fields from Eqs. \eqref{eq:firstField}--\eqref{eq:lastField} as 
\begin{align}
&\boldsymbol{\mathcal{D}}_{uk}^{\mathrm{in/out},\lambda}(\boldsymbol{r}) = \sqrt{\frac{\hbar\omega_u}{4\pi}} \boldsymbol{\mathsf{d}}^{(\ell)}_u(\boldsymbol{r})\tilde{{\mathcal{D}}}_{uk}^{\mathrm{in/out},\lambda}e^{i\kappa^{(\ell)}_u\zeta} \,\,\,\,[\boldsymbol{r}\in \, \mathrm{ring}\, \ell],
\end{align}
and we can write
\begin{equation}
\begin{split}
    K_1^{\lambda\lambda'}(k_1,...,k_4) 
    &=\frac{\hbar^2\sqrt{\omega_{S_{1}}\omega_{I}}v_{P_{1}}v_{P_{1}} \mathcal{L}^{(1)}\gamma_{\mathrm{NL}}^{(1)}}{4\pi^2} \left[ \tilde{{\mathcal{D}}}_{S_{1},k_1}^{\mathrm{out},\lambda} \tilde{{\mathcal{D}}}_{I,k_2}^{\mathrm{out},\lambda'}\right]^* \tilde{{\mathcal{D}}}_{P_{1},k_3}^{\mathrm{in,ac}} \tilde{{\mathcal{D}}}_{P_{1},k_4}^{\mathrm{in,ac}}
    \end{split}
    \label{eq:k1Work}
\end{equation}

\begin{equation}
\begin{split}
   K_2^{\mu\mu'\mu''}(k_1,...,k_4) 
    &=  \frac{\hbar^2\sqrt{\omega_{S_{2}}\omega_{S_{3}}}v_{I}v_{P_{2}} \mathcal{L}^{(2)}\gamma_{\mathrm{NL}}^{(2)}}{2\pi^2} \left[ \tilde{{\mathcal{D}}}_{S_{2},k_1}^{\mathrm{out},\mu} \tilde{{\mathcal{D}}}_{S_{3},k_2}^{\mathrm{out},\mu'}\right]^* \tilde{{\mathcal{D}}}_{I,k_3}^{\mathrm{out},\mu''} \tilde{{\mathcal{D}}}_{P_{2},k_4}^{\mathrm{in,ac}},
    \end{split}
    \label{eq:k2Work}
\end{equation}
where we define 
\begin{align}
    \gamma_{\mathrm{NL}}^{(1)} & \equiv \frac{3\sqrt{\omega_{P_{1}}\omega_{P_{1}}}}{4\epsilon_0v_{P_{1}}v_{P_{1}} \mathcal{L}^{(1)}} \int \mathrm{d}\boldsymbol{r}  \Gamma_{ijkl}^{(3)} (\boldsymbol{r}) \left[ {\mathsf{d}}^{(1),i}_{S_{1}}(\boldsymbol{r})
     {\mathsf{d}}^{(1),j}_{I}(\boldsymbol{r})\right]^{*}  {\mathsf{d}}^{(1),k}_{P_{1}}(\boldsymbol{r}) {\mathsf{d}}^{(1),l}_{P_{1}}(\boldsymbol{r})e^{i(\kappa^{(1)}_{P_{1}} + \kappa^{(1)}_{P_{1}} - \kappa^{(1)}_{I} - \kappa^{(1)}_{S_{1}})\zeta},\\
    \gamma_{\mathrm{NL}}^{(2)} & \equiv \frac{3\sqrt{\omega_{I}\omega_{P_{2}}}}{4\epsilon_0v_{I}v_{P_{2}} \mathcal{L}^{(2)}} \int \mathrm{d}\boldsymbol{r}  \Gamma_{ijkl}^{(3)} (\boldsymbol{r}) \left[ {\mathsf{d}}^{(2),i}_{S_{2}}(\boldsymbol{r})
    {\mathsf{d}}^{(2),j}_{S_{3}}(\boldsymbol{r})\right]^{*}  {\mathsf{d}}^{(2),k}_{I}(\boldsymbol{r}) {\mathsf{d}}^{(2),l}_{P_{2}}(\boldsymbol{r})e^{i(\kappa^{(2)}_{I} + \kappa^{(2)}_{P_{2}} - \kappa^{(2)}_{S_{3}} - \kappa^{(2)}_{S_{2}})\zeta}.
\end{align}
Again, we note that to arrive to such expressions, full permutation symmetry of the $\Gamma^{(3)}$ tensor is assumed, which results from assuming that the indices of the second- and third-order nonlinear susceptibilities can be permuted freely \cite{Quesada:22}. For the materials of interest in this work, the contribution of the second-order nonlinear susceptibility to the $\Gamma^{(3)}$ tensor is negligible, thus we can write the $\Gamma^{(3)}$ tensor components as \cite{Quesada:22}:
\begin{equation}
\Gamma^{(3)}_{ijkl}(\boldsymbol{r}) = \frac{\chi^{(3)}_{ijkl}(\boldsymbol{r})}{\epsilon_0^2\varepsilon_1^4(\boldsymbol{r})}.
    \label{eq:Tensors}
\end{equation}
In identifying the effective nonlinear coefficient we neglect any frequency dependence on the dielectric constants, since we consider frequencies in the same range \cite{Quesada:22}. The spatial integral is restricted to the volume of the ring: We assume that the nonlinear interaction occurs only inside the ring, due to the field enhancement there compared to the waveguide, and since the nonlinear susceptibilities of the SiO$_2$ cladding are orders of magnitude smaller than those of the material that constitutes the ring. We can relate the linear displacement fields to the electric fields
\begin{equation} 
    \bm{\mathsf{d}}^{(\ell)}_u(\boldsymbol{r}) = \epsilon_0 \varepsilon_1(\boldsymbol{r}) \bm{\mathsf{e}}^{(\ell)}_u(\boldsymbol{r}_\perp,\zeta), 
    \label{eq:Electric_Displacement_Small}
\end{equation}
where $\boldsymbol{r}_\perp$ is a vector in the cross-section perpendicular to the propagation direction of the field $\zeta$, and we can write
\begin{equation}
\begin{split}
    \gamma_{\mathrm{NL}}^{(\ell)} & = \frac{3\epsilon_0\sqrt{\omega_{{u_{3}}}\omega_{{u_{4}}}}}{4v_{{u_{3}}}v_{{u_{4}}} \mathcal{L}^{(\ell)}} \int \mathrm{d}\boldsymbol{r}_\perp \mathrm{d}\zeta \chi^{(3)}_{ijkl}(\boldsymbol{r}) \left[   {\mathsf{e}}^{(\ell),i}_{{u_{1}}}(\boldsymbol{r}_{\perp})
       {\mathsf{e}}^{(\ell),j}_{{u_{2}}}(\boldsymbol{r}_{\perp})\right]^{*}    {\mathsf{e}}^{(\ell),k}_{{u_{3}}}(\boldsymbol{r}_{\perp})  {\mathsf{e}}^{(\ell),l}_{{u_{4}}}(\boldsymbol{r}_{\perp})e^{i(\kappa^{(\ell)}_{{u_{3}}} + \kappa^{(\ell)}_{{u_{4}}} - \kappa^{(\ell)}_{{u_{2}}} - \kappa^{(\ell)}_{{u_{1}}})\zeta},
    \label{eq:GammaEs}
    \end{split}
\end{equation}
where the components of $\boldsymbol{{\mathsf{e}}}^{(\ell)}_{u}(\boldsymbol{r}_{\perp})$ can be extracted from an eigenmode solver for the specific structure and the integral over $\boldsymbol{r}_\perp$ can be evaluated numerically. 

We have dropped the $\zeta$ dependence on the electric fields, writing only $\boldsymbol{\mathsf{e}}^{(\ell)}_u(\boldsymbol{r}_\perp)$. This implicitly implies that the components of the fields $\boldsymbol{\mathsf{e}}^{(\ell)}_u(\boldsymbol{r}_\perp)$ and of the tensor $\chi^{(3)}_{ijkl}$ are in the frame rotating with the light's propagation along the rings: For each ring, we define this as the ring frame with coordinates $xyz$ (see Fig. \ref{fig:rings_modes}a). In the ring frame, a vector $\boldsymbol{r}_\perp$ lies in the $xz$ plane, and $\zeta=\phi R_{\ell}$ goes along the $y$ axis. The integral over $\boldsymbol{r}_\perp$ ranges over the cross-section of the ring, and the one over $\zeta$ ranges from $0$ to $\mathcal{L}^{(\ell)}$. For the interactions of interest, we consider processes with phase matched wavenumbers $\kappa^{(\ell)}_{{u_{3}}} + \kappa^{(\ell)}_{{u_{4}}} - \kappa^{(\ell)}_{{u_{2}}} - \kappa^{(\ell)}_{{u_{1}}} = 0$; thus we can write
\begin{equation}
\begin{split}
    \gamma_{\mathrm{NL}}^{(\ell)} & = \frac{3\epsilon_0\sqrt{\omega_{{u_{3}}}\omega_{{u_{4}}}}}{4v_{{u_{3}}}v_{{u_{4}}} \mathcal{L}^{(\ell)}} \int \mathrm{d}\boldsymbol{r}_\perp \mathrm{d}\zeta \chi^{(3)}_{ijkl}(\boldsymbol{r}) \left[   {\mathsf{e}}^{(\ell),i}_{{u_{1}}}(\boldsymbol{r}_{\perp})
       {\mathsf{e}}^{(\ell),j}_{{u_{2}}}(\boldsymbol{r}_{\perp})\right]^{*}    {\mathsf{e}}^{(\ell),k}_{{u_{3}}}(\boldsymbol{r}_{\perp})  {\mathsf{e}}^{(\ell),l}_{{u_{4}}}(\boldsymbol{r}_{\perp}). 
       \label{eq:GammaEs2}
    \end{split}
\end{equation}

The III-V semiconductors of interest, such as AlGaAs, belong to the $\bar{4}3m$ point group and have cubic symmetry \cite{bib:Boyd}. In general, for crystals in this point group the third-order nonlinear susceptibility $\chi^{(3)}$ tensor has 21 nonzero components, of which 4 are independent. We introduce the lab frame (axes $x'y'z'$), which coincides with the principal axes of the crystal, and the nonzero tensor components of the $\chi^{(3)}$ tensor are (\cite{bib:Boyd}):
\begin{align}
    & \chi^{(3)}_{x'x'x'x'}=\chi^{(3)}_{y'y'y'y'}=\chi^{(3)}_{z'z'z'z'}, \\
    & \chi^{(3)}_{x'x'y'y'}=\chi^{(3)}_{y'y'x'x'} =\chi^{(3)}_{y'y'z'z'}=\chi^{(3)}_{z'z'y'y'}=\chi^{(3)}_{z'z'x'x'}=\chi^{(3)}_{x'x'z'z'}, \label{eq:Chi3FullFirst}\\
    & \chi^{(3)}_{x'y'x'y'}=\chi^{(3)}_{y'x'y'x'}=\chi^{(3)}_{y'z'y'z'}=\chi^{(3)}_{z'y'z'y'}=\chi^{(3)}_{z'x'z'x'}=\chi^{(3)}_{x'z'x'z'} , \\
    & \chi^{(3)}_{x'y'y'x'}=\chi^{(3)}_{y'x'x'y'}=\chi^{(3)}_{y'z'z'y'}=\chi^{(3)}_{z'y'y'z'}=\chi^{(3)}_{z'x'x'z'}=\chi^{(3)}_{x'z'z'x'} . \label{eq:Chi3FullLast}
\end{align}
The components can be written as (\cite{AfsharV.:09,Johnson2019FullVectorialAlGaAs})
\begin{equation}
\chi^{(3)}_{i'j'k'l'} = \chi^{(3)}_{x'x'y'y'}\delta_{i'j'}\delta_{k'l'} + \chi^{(3)}_{x'y'x'y'}\delta_{i'k'}\delta_{j'l'} + \chi^{(3)}_{x'y'y'x'}\delta_{i'l'}\delta_{j'k'} + (\chi^{(3)}_{x'x'x'x'} - \chi^{(3)}_{x'x'y'y'} - \chi^{(3)}_{x'y'x'y'}- \chi^{(3)}_{x'y'y'x'})\delta_{i'j'k'l'},
    \label{eq:tensorwithdelta}
\end{equation}
where $\delta_{ab}$ is the usual Kronecker delta function, and $\delta_{abcd}$ is equal to unity if $a=b=c=d$, and zero otherwise. 

In arriving at expressions for the nonlinear parameters such as Eq. \eqref{eq:GammaEs}, permutation symmetry of the indices of the $\Gamma^{(3)}$ tensor was assumed, and for AlGaAs this is analogous to assuming Kleinman symmetry. Under this approximation, Eqs. \eqref{eq:Chi3FullFirst} to \eqref{eq:Chi3FullLast} are all equal, which implies that the third-order nonlinear susceptibility tensor has only two independent components: $\chi^{(3)}_{x'x'x'x'}$ and $\chi^{(3)}_{x'x'y'y'}$. Thus, under Kleinman symmetry, we can write the tensor components as 
\begin{equation}
    \chi^{(3)}_{i'j'k'l'} = \chi^{(3)}_{x'x'y'y'}(\delta_{i'j'}\delta_{k'l'} + \delta_{i'k'}\delta_{j'l'} + \delta_{i'l'}\delta_{j'k'}) + (\chi^{(3)}_{x'x'x'x'} - 3\chi^{(3)}_{x'x'y'y'})\delta_{i'j'k'l'}.
    \label{eq:tensorwithdelta_klein}
\end{equation}
In this framework, the third-order nonlinear optical interactions can be described by two independent components of the nonlinear susceptibility: $\chi^{(3)}_{x'x'x'x'}$ and $\chi^{(3)}_{x'x'y'y'}$. As mentioned in Johnson \cite{Johnson2019FullVectorialAlGaAs}, assuming Kleinman symmetry for frequencies close to the half band gap (such as those considered in this work) often leads to underestimating the nonlinear interaction strength; thus we anticipate that our calculated results (e.g., generation rates) to be slightly lower than those obtainable from experiments in fabricated samples.

From the form that we have for the nonlinear parameters in Eq. \eqref{eq:GammaEs}, to contract the indices using the electric field mode profiles in the ring frame we need the tensor components of $\chi^{(3)}$ in the ring frame. For a vector $\boldsymbol{v}$, with components $v_i$ and $v_{i'}$ in the ring and lab frames, respectively, we can relate its components according to
\begin{equation}
\begin{split}
v_i&=\boldsymbol{M}_\phi^{i{i'}}v_{i'},\\
v_{i'}&=\boldsymbol{M}_\phi^{ii'}v_i,
\end{split}
\label{eq:v_transform}
\end{equation}
where 
\begin{equation}
\boldsymbol{M}_\phi=\begin{bmatrix}\cos\phi&\sin\phi&0\\-\sin\phi&\cos\phi&0\\0&0&1\end{bmatrix}.
\end{equation}
It follows that for a rank-$n$ tensor, we can identify its components in the ring frame with
\begin{equation}
    \Upsilon^{(n)}_{i_1,\dots,i_n} = \boldsymbol{M}_\phi^{i_1{i_1'}}\dots\boldsymbol{M}_\phi^{i_n{i_n'}} \Upsilon^{(n)}_{i'_1,\dots,i'_n}.
\end{equation}
To simplify the discussion, we specialize to the case considered in the text where all four fields are TE$_{00}$ polarized. This implies that to good approximation the main component of the fields are $\mathsf{e}^{x}_{u}$, and all of the terms involving $\mathsf{e}^{y}_{u}$ and $\mathsf{e}^{z}_{u}$ are near zero. In this case, only the contraction of the fields in the ring frame with the $\chi^{(3)}_{xxxx}$ tensor component contribute to the integral for the nonlinear parameters from Eq. \eqref{eq:GammaEs2}. To good approximation we can write
\begin{equation}
    \begin{split}
        \chi^{(3)}_{ijkl}(\boldsymbol{r})\left[   {\mathsf{e}}^{i}_{{u_{1}}}
       {\mathsf{e}}^{j}_{{u_{2}}}\right]^{*}    {\mathsf{e}}^{k}_{{u_{3}}}  {\mathsf{e}}^{l}_{{u_{4}}} \approx \left[ {\mathsf{e}}^{x}_{{u_{1}}} {\mathsf{e}}^{x}_{{u_{2}}}\right]^{*}    {\mathsf{e}}^{x}_{{u_{3}}}  {\mathsf{e}}^{x}_{{u_{4}}}  \times \left( A - \frac{F}{2}\sin^2(2\phi)\right),
    \end{split}
    \label{eq:FullFieldsTE_1}
\end{equation}
where we defined
\begin{align}
    A&\equiv\chi^{(3)}_{x'x'x'x'},\\
    B&\equiv\chi^{(3)}_{x'x'y'y'},\\
    F&\equiv A-3B.\label{eq:Fdef}
\end{align}
The effect of the anisotropy on the nonlinear parameter is described by the term with $F$, where if $F>0$ the anisotropy would reduce the nonlinear parameter and if $F<0$ it would increase the nonlinear parameter, since $\int\mathrm{d}\zeta \sin^2(2\phi)>0$. 

{The magnitude of $F$, and hence the effect of the anisotropy, depends on the material considered. Experimental work and modelling (without assuming Kleinman symmetry) was done for AlGaAs with 18\% aluminum content \cite{Hutchings:95,Hutchings:97,Johnson2019FullVectorialAlGaAs}, and we assume that the AlGaAs we consider in this manuscript (30\% aluminum content) has similar properties. When the frequencies of the generated light are close to each other, as is the case in this work, to good approximation the tensor components $\chi^{(3)}_{x'y'x'y'}$ and $ \chi^{(3)}_{x'y'x'y'}$ can be set to be equal, in which case one can introduce \cite{Hutchings:95,Hutchings:97,Johnson2019FullVectorialAlGaAs} the anisotropy parameter $\sigma=(\chi^{(3)}_{x'x'x'x'}-\chi^{(3)}_{x'x'y'y'}-2\chi^{(3)}_{x'y'x'y'})/(2\chi^{(3)}_{x'x'x'x'})$, which characterizes the effect of the anisotropy similar to the $F$ we introduced in Eq. \eqref{eq:Fdef}. In the cited studies, the values of $\sigma$ are negative (and hence $F<0$), implying that the anisotropy of the material effectively increases the nonlinear parameters.} For this manuscript, to ensure that we do not overestimate the nonlinear parameters, we use $F=0$; we expect the nonlinear parameters to be slightly larger than those we compute. A more thorough investigation of the change of the magnitudes of the nonlinear parameters from the anisotropy could be done, although significant enhancements of the nonlinear parameters are not expected. 

Thus, for the four-wave mixing processes in AlGaAs that we consider in this manuscript, we can compute the nonlinear coefficients to good approximation with
\begin{equation}
\begin{split}
    \gamma_{\mathrm{NL}}^{(\ell)} & \approx  \frac{3\chi^{(3)}_{x'x'x'x'} \epsilon_0\sqrt{\omega_{{u_{3}}}\omega_{{u_{4}}}}}{4v_{{u_{3}}}v_{{u_{4}}}} \int \mathrm{d}\boldsymbol{r}_\perp \left[   {\mathsf{e}}^{(\ell),x}_{{u_{1}}}(\boldsymbol{r}_{\perp})
       {\mathsf{e}}^{(\ell),x}_{{u_{2}}}(\boldsymbol{r}_{\perp})\right]^{*}    {\mathsf{e}}^{(\ell),x}_{{u_{3}}}(\boldsymbol{r}_{\perp})  {\mathsf{e}}^{(\ell),x}_{{u_{4}}}(\boldsymbol{r}_{\perp}). 
       \label{eq:GammaEs3}
    \end{split}
\end{equation}

}
\section{Asymptotic fields for dual rings coupled to a waveguide\label{supp:Fields}}

\begin{figure}[ht]
    \centering
\includegraphics[width=0.75\linewidth]{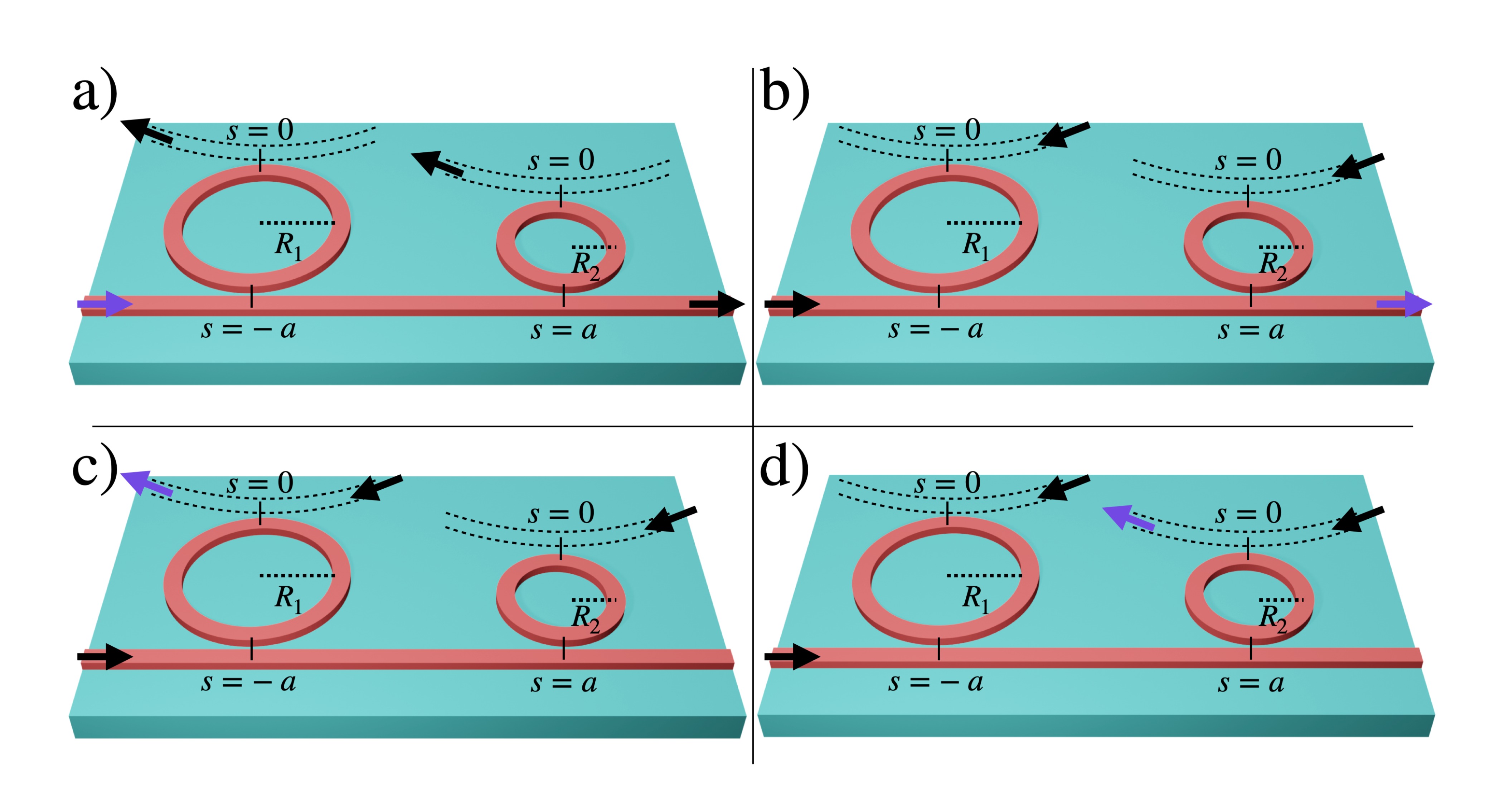}
    \caption{Schematic of the asymptotic-in state for the actual waveguide (a), the asymptotic-out for the actual waveguide (b), the asymptotic-out for the first phantom waveguide (c), and the asymptotic-out for the second phantom waveguide (d).}
    \label{fig:asyStates}
\end{figure}

The asymptotic-in and -out mode fields we consider form a complete set for the structures of interest, and we can write the total displacement field
\begin{equation}
    \boldsymbol{D}(\boldsymbol{r}) = \sum_u \boldsymbol{D}_{u}^{\mathrm{in/out}}(\boldsymbol{r}) + \mathrm{H.c.},
    \label{eq:TotalAsyFields1}
\end{equation}
with 
\begin{align}
    \boldsymbol{D}_u^{\mathrm{in/out}}(\boldsymbol{r}) = \sum_{\lambda} \int \mathrm{d}k \boldsymbol{\mathcal{D}}_{uk}^{\mathrm{in/out}, \lambda}(\boldsymbol{r}) a_u^{\mathrm{in/out},\lambda}(k) + \mathrm{H.c..} \label{eq:Dasy_def2}
\end{align} 
We provide the fields inside the microring resonators for each case in Fig. 4, obtained from the Heisenberg equations for the specific asymptotic state \cite{PhysRevA.106.043707}. Using $\boldsymbol{r}_\perp$ to denote a vector along the cross-section of the waveguide constructing the ring, and $\zeta$ to identify the azimuthal coordinate along the circumference of the ring (see Fig. \ref{fig:rings_modes}a), the asymptotic-in fields for the actual waveguide (Fig. \ref{fig:asyStates}a) are 
\begin{align}
&\boldsymbol{\mathcal{D}}_{uk}^{\mathrm{in,ac}}(\boldsymbol{r}) = -\sqrt{\frac{\hbar\omega_u}{4\pi}} \boldsymbol{\mathsf{d}}^{(1)}_u(\boldsymbol{r}_\perp)F_{u-}^{\mathrm{ac}}(k) e^{-i(k-K_u)a}e^{i\kappa^{(1)}_u\zeta},\nonumber\\
&[\boldsymbol{r}\in \, \mathrm{ring\, 1}] \label{eq:firstField}\\
&\boldsymbol{\mathcal{D}}_{uk}^{\mathrm{in,ac}}(\boldsymbol{r}) = -\sqrt{\frac{\hbar\omega_u}{4\pi}} \boldsymbol{\mathsf{d}}^{(2)}_u(\boldsymbol{r}_\perp)F_{u-}^{\mathrm{ac}(2)}(k) \left( 1 + \frac{i\gamma_u^{\mathrm{ac}(1)}}{v_u^{\mathrm{ac}}}\sqrt{\mathcal{L}^{(1)}} F_{u-}^{\mathrm{ac}(1)}(k)   \right) e^{i(k-K_u)a}e^{i\kappa^{(2)}_u\zeta},\nonumber\\
&[\boldsymbol{r}\in \, \mathrm{ring\, 2}]
\end{align}
the asymptotic-out fields for the actual waveguide (Fig. \ref{fig:asyStates}b) are 
\begin{align}
&\boldsymbol{\mathcal{D}}_{uk}^{\mathrm{out,ac}}(\boldsymbol{r}) = -\sqrt{\frac{\hbar\omega_u}{4\pi}} \boldsymbol{\mathsf{d}}^{(1)}_u(\boldsymbol{r}_\perp)F_{u+}^{\mathrm{ac}(1)}(k) \left( 1 - \frac{i\gamma_u^{\mathrm{ac}(2)}}{v_u^{\mathrm{ac}}}\sqrt{\mathcal{L}^{(2)}}  F_{u+}^{\mathrm{ac}(2)}(k)  \right)  e^{-i(k-K_u)a} e^{i\kappa^{(1)}_u\zeta},\nonumber\\
&[\boldsymbol{r}\in \, \mathrm{ring\, 1}]\\
&\boldsymbol{\mathcal{D}}_{uk}^{\mathrm{out,ac}}(\boldsymbol{r}) = -\sqrt{\frac{\hbar\omega_u}{4\pi}} \boldsymbol{\mathsf{d}}^{(2)}_u(\boldsymbol{r}_\perp) F_{u+}^{\mathrm{ac}(2)}(k)  e^{i(k-K_u)a}   e^{i\kappa^{(2)}_u\zeta},\nonumber\\
&[\boldsymbol{r}\in \, \mathrm{ring\, 2}]
\end{align}
the asymptotic-out fields for the first phantom waveguide (Fig. \ref{fig:asyStates} c)) are
\begin{align}
&\boldsymbol{\mathcal{D}}_{uk}^{\mathrm{out,ph}(1)}(\boldsymbol{r}) =-\sqrt{\frac{\hbar\omega_u}{4\pi}} \boldsymbol{\mathsf{d}}^{(1)}_u(\boldsymbol{r}_\perp) F_{u+}^{\mathrm{ph}(1)}(k)  e^{-i(k-K_u)a} e^{i\kappa^{(1)}_u\zeta}, \nonumber\\
&[\boldsymbol{r}\in \, \mathrm{ring\, 1}]\\
&\boldsymbol{\mathcal{D}}_{uk}^{\mathrm{out,ph}(1)}(\boldsymbol{r}_\perp) = 0,\nonumber\\
&[\boldsymbol{r}\in \, \mathrm{ring\, 2}]
\end{align}
and the asymptotic-out fields for the second phantom waveguide (Fig. \ref{fig:asyStates}d) are 
\begin{align}
&\boldsymbol{\mathcal{D}}_{uk}^{\mathrm{out,ph}(2)}(\boldsymbol{r}) =-\sqrt{\frac{\hbar\omega_u}{4\pi}} \boldsymbol{\mathsf{d}}^{(1)}_u(\boldsymbol{r}_\perp) F_{u+}^{\mathrm{ac}(1)}(k) \left(  -\frac{i\gamma_u^{\mathrm{ac}(2)}}{v_u^{\mathrm{ac}}}\sqrt{\mathcal{L}^{(2)}}F_{u+}^{\mathrm{ph}(2)}(k)  \right)  e^{-i(k-K_u)a} e^{i\kappa^{(1)}_u\zeta} \nonumber\\
&[\boldsymbol{r}\in \, \mathrm{ring\, 1}]\\
&\boldsymbol{\mathcal{D}}_{uk}^{\mathrm{out,ph}(2)}(\boldsymbol{r}) = -\sqrt{\frac{\hbar\omega_u}{4\pi}} \boldsymbol{\mathsf{d}}^{(2)}_u(\boldsymbol{r}_\perp) F_{u+}^{\mathrm{ph}(2)}(k)  e^{i(k-K_u)a} e^{i\kappa^{(2)}_u\zeta}, \nonumber\\
&[\boldsymbol{r}\in \, \mathrm{ring\, 2}]\label{eq:lastField}
\end{align}
where $\boldsymbol{\mathsf{d}}^{(\ell)}_u(\boldsymbol{r})$ are the spatial mode profiles in the $\ell^{\mathrm{th}}$ ring ($\ell\in\{1,2\}$) associated with the frequency range $u$ \cite{Quesada:22}. We have also introduced a field enhancement factor
\begin{equation}
     {F}_{u \pm}^{ \lambda(\ell)} (k) = \frac{1}{\sqrt{\mathcal{L}^{(\ell)}}} \left( \frac{({\gamma}_u^{\lambda(\ell)})^*}{v_u(K_u - k)\pm i\bar{\Gamma}_u^{(\ell)}} \right).
    \label{eq:FieldEnchance_Definition2}
\end{equation}
The spatial modes are normalized according to 
\begin{equation}
\begin{split}
    \int \frac{v_p(\boldsymbol{r}_\perp;\omega_{u})}{v_g(\boldsymbol{r}_\perp;\omega_{u})} \frac{\bm{\mathsf{d}}_{u}^{(\ell)*} (\boldsymbol{r}_\perp) \cdot \bm{\mathsf{d}}_{u}^{(\ell)} (\boldsymbol{r}_\perp)}{\epsilon_0 \varepsilon_1(\boldsymbol{r}_\perp;\omega_u)}  \mathrm{d} \boldsymbol{r}_\perp = 1,
\end{split}
    \label{eq:NormalizationConditionDfieldsNEW}
\end{equation}
where $v_p$ and $v_g$ are the local phase and group velocities of the material \cite{Quesada:22}. 

\end{document}